\documentclass[twocolumn,aps,prl,showpacs,amsmath,superscriptaddress,longbibliography,notitlepage]{revtex4-2}
\usepackage{amssymb}
\usepackage{braket}   
\usepackage{amsmath}
\usepackage{mathrsfs}
\usepackage{graphicx} 
\usepackage{float}  
\usepackage[caption=false]{subfig} 
\usepackage[pdftex,colorlinks,linkcolor={red!75!black},citecolor={red!75!black},urlcolor={blue!75!black}]{hyperref}
\usepackage{verbatim} 
\usepackage{epstopdf}
\usepackage[normalem]{ulem}
\usepackage{xcolor} 
\usepackage{bm}     
\usepackage{soul}
\usepackage{ulem} 
\usepackage[flushleft]{threeparttable}

\definecolor{mypurple}{RGB}{128,0,128} 
\definecolor{myorange}{RGB}{255,140,0}  
\definecolor{myblue}{RGB}{0,160,233}  
\definecolor{mygreen}{RGB}{34,172,56}

\definecolor{colorRhoA}{HTML}{2A9D8F}   
\definecolor{colorRhoB}{HTML}{E9C46A}   

\begin{document} 
\title{Anyon-induced non-Hermitian topological phases}
\author{Yi-An Wang}
\affiliation{Quantum Science Center of Guangdong-Hong Kong-Macao Greater Bay Area (Guangdong), Shenzhen, 518045, China}
\affiliation{Department of Physics, State Key Laboratory of Surface Physics, and Key Laboratory of Micro and Nano Photonic Structures (Ministry of Education), Fudan University, Shanghai 200438, China}

\author{Kun Ding}\email{kunding@fudan.edu.cn}
\affiliation{Department of Physics, State Key Laboratory of Surface Physics, and Key Laboratory of Micro and Nano Photonic Structures (Ministry of Education), Fudan University, Shanghai 200438, China}

\author{Linhu Li}\email{lilinhu@quantumsc.cn}
\affiliation{Quantum Science Center of Guangdong-Hong Kong-Macao Greater Bay Area (Guangdong), Shenzhen, 518045, China}
\begin{abstract} 
We show that anyonic exchange statistics can activate non-Hermitian point-gap topology in models that are topologically trivial in its absence. The emergent topology oscillates more rapidly with the statistical phase as the anyon number increases, and exhibits a parity dependence on the particle number. A perturbative analysis reveals the mechanism: fractional statistics induces a mismatch between momentum terms that, combined with sublattice-dependent dissipation, produces particle-number-dependent non-reciprocity and complex spectral winding. As these effects rely on the formation and exchange of interaction-bound anyons, our results establish exchange statistics as a resource for enabling non-Hermitian topology under programmed dissipation.
\end{abstract}

\maketitle
\emph{Introduction---}
Non-Hermitian physics profoundly reshapes the notion of topological phases of matter~\cite{xiong2018does,PhysRevLett.120.146402,PhysRevA.97.052115,li2019geometric,yao2018edge,yao2018non,PhysRevLett.126.216405,li2021quantized,
gong2018topological,PhysRevX.9.041015,PhysRevLett.124.086801,PhysRevLett.125.126402,borgnia2020non,
liu2019second,PhysRevLett.123.016805,PhysRevB.106.035425,PhysRevLett.128.223903,
PhysRevLett.126.010401,PhysRevResearch.4.L022064,
lee2019topological,PhysRevLett.133.266604,
bergholtz2021exceptional,PhysRevLett.127.034301,guo2023exceptional,xue2026essay,
helbig2020generalized,PhysRevLett.126.215302,wang2021detecting,lin2022topological,lin2022observation,wang2021generating,wang2021topological,
ding2022non,okuma2023non,lin2023topological}. 
In particular, the complex energy spectrum enables spectral winding topology protected by a point gap under periodic boundary conditions (PBCs),
which gives rise to the non-Hermitian skin effect (NHSE) under open boundary conditions (OBCs), characterized by the non-reciprocal accumulation of eigenstates at system boundaries~\cite{yao2018edge,PhysRevX.9.041015,PhysRevLett.124.086801,PhysRevLett.125.126402}. 
In the many-body regime, quantum statistics and particle interactions further enrich the NHSE landscape, leading to emergent phenomena such as real-space Fermi surfaces and  boundary condensation~\cite{mu2020emergent,cao2023many,garbe2024bosonic,mao2023non}, and distinct skin localization across internal subspaces~\cite{lee2021many,shen2022non,faugno2022interaction,kawabata2022many,li2023non,yoshida2024non,qin2024occupation,shen2024enhanced,gliozzi2024many,PhysRevLett.133.136502,wang2025non,hu2025many,qin2026manybody,lei2026inter}.

Anyons, characterized by fractional exchange statistics beyond bosons and fermions, have attracted intense interest as the origin of exotic quantum phases such as the fractional quantum Hall effect~\cite{leinaas1977theory,wilczek1982magnetic,tsui1982two,laughlin1983anomalous,halperin1984statistics,arovas1984fractional,kitaev2006anyons,PhysRevLett.99.247203,bauer2014chiral,goldin2023prediction}, and as a foundation for fault-tolerant quantum computing~\cite{kitaev2003fault,das2005topologically,nayak2008non,carrega2021anyons,iqbal2024non,lee2023partitioning,lee2015geometric,PhysRevLett.121.237401}.
The study of two-dimensional anyons has long been intertwined with conformal field theories and quantum groups, which have inspired intriguing developments in non-Hermitian physics~\cite{pasquier1990common,feiguin2007interacting,trebst2008collective,freedman2012galois}.
On the other hand, Abelian anyonic behaviors have also been predicted in one dimension (1D)~\cite{PhysRevLett.83.1275,hao2008ground,hao2009ground,keilmann2011statistically,hao2012dynamical,PhysRevLett.115.053002,PhysRevA.94.013611,PhysRevLett.118.120401,PhysRevA.97.023631,PhysRevLett.121.250404,PhysRevLett.121.250404}  and recently realized in ultracold atomic systems with a continuously tunable statistical angle~\cite{kwan2024realization,dhar2025observing}. This raises the question of what role such tunable statistics can play in non-Hermitian topology. To date, anyonic statistics has been shown only to tune a pre-existing non-Hermitian topology, for instance, suppressing the NHSE~\cite{qin2025dynamical} or modifying criticality between skin modes~\cite{qin2026anyon}. Whether fractional statistics can themselves activate non-Hermitian topology in an otherwise trivial system, has remained an open question.

\begin{figure} 
\centering
\includegraphics[width=\linewidth]{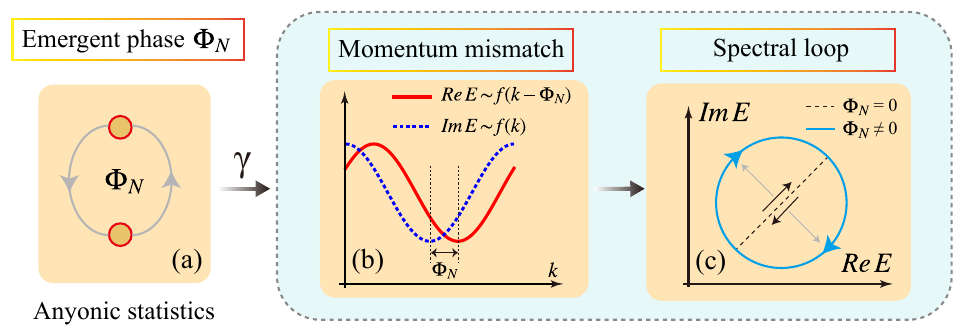}  
\caption{Schematic mechanism of the anyon-induced non-Hermitian topology. 
        (a) Fractional exchange statistics endows an $N$-particle bound state with a many-body statistical phase $\Phi_N=N(N-1)\theta/2$. 
        (b) With non-Hermiticity ($\gamma$ in our explicit model below),  $\Phi_N$ affects the real and imaginary energies of the bound state differently, so that ${\rm Re} E$ (red line) and ${\rm Im} E$ (blue line) become misaligned as $k$ varies. (c) This $\Phi_N$-induced Re-Im misalignment transforms the spectrum into a closed spectral loop (opening a point gap). The black and cyan arrows show the energy trajectory for $\Phi_N=0$ and $\Phi_N\neq0$, respectively, as $k$ varies.
} 
\label{fig:1}
\end{figure}

In this paper, we unveil that anyonic statistics can activate nontrivial non-Hermitian point-gap topology, and its corresponding NHSE of multi-particle bound states, in a dissipative non-Hermitian ladder model.
Namely, the anyonic statistical angle generates a momentum mismatch between different Hamiltonian terms, which, when interacting with sublattice-dependent dissipation, induces a non-trivial point gap that generates non-reciprocal pumping to the anyons (See Fig.~\ref{fig:1} for a schematic illustration).
The phenomenon and underlying mechanism are captured by a projected model in the bound-state subspace, 
where non-reciprocity arises from a phase factor depending on both the statistical phase $\theta$ and the particle number $N$.
Consequently, the emergent point-gap topology exhibits unique behaviors, including its rapid oscillation with $\theta$ for large $N$ and parity dependence on $N$.

\begin{figure} 
\centering
\includegraphics[width=\linewidth]{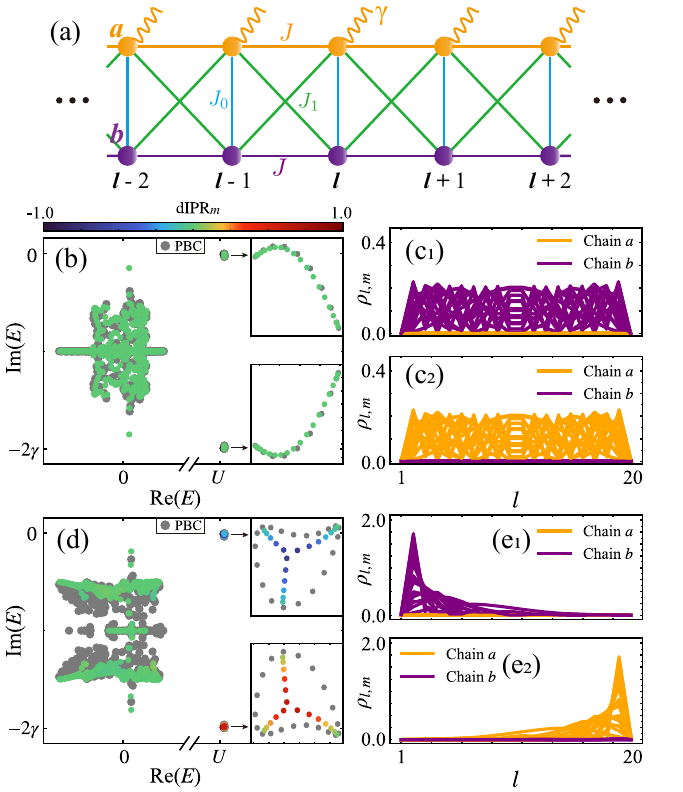}  
\caption{
Emergence of the anyonic NHSE in the two-particle bound-state clusters. 
(a) Schematic of the non-Hermitian ladder model. 
(b) Complex spectra for $N=2$ and $\theta=0$ (bosons), 
with ${\rm dIPR}_m\approx0$ for all eigenstates.
The PBC and OBC spectra for bound states with ${\rm Re}(E)\approx U$ form nearly identical curves.
(c) State distributions in the two bound-state clusters, which are both extended.
(d) and (e) the same as (b) and (c), but with $\theta=\pi/2$ (anyons).
The bound-state clusters develop loop spectra under PBCs [after applying the transformation in Eq.~\eqref{eq:JW}], and opposite skin localizations under OBCs.
Other parameters are $L=20$, $N=2$, $J_0=J_1=3$, $J=1$, $\gamma=2$, and $U=100$.
} 
\label{fig:2}
\end{figure}

\emph{Model and anyonic NHSE.---}
We consider a 1D non-Hermitian two-chain ladder lattice loaded with Abelian anyons, 
described by the Hamiltonian:
{\small
\begin{align}
\hat{H}_{\rm{a}} = 
& - \sum_{l=1}^L \Bigg[
    \sum_{\beta=a,b} J \hat{\beta}_l^\dagger \hat{\beta}_{l+1}  
    + J_1 \hat{a}_l^\dagger \hat{b}_{l+1} + J_1 \hat{b}_l^\dagger \hat{a}_{l+1}  
    + J_0 \hat{a}_l^\dagger \hat{b}_l
\Bigg] \notag \\
& + \text{h.c.}
+ \frac{U}{2} \sum_{\beta=a,b} \sum_{l=1}^L 
    \hat{n}_{\beta,l} \left( \hat{n}_{\beta,l} - 1 \right) 
- i \gamma \sum_{l=1}^L \hat{a}_l^\dagger \hat{a}_l,\label{eq:H_creutz_anyon}
\end{align}
}%
as illustrated in Fig. \ref{fig:2}(a).  
$J$, $J_0$, and $J_1$ are Hermitian hopping amplitudes, which are assumed to be positive hereafter.
$U$ denotes a Hubbard interaction, and $\gamma$ represents the non-Hermitian dissipation on chain $a$. 
$\hat{\beta}$ ($\hat{\beta}^\dagger$) are the anyon annihilation (creation) operators on chain $\beta$, 
satisfying
\begin{subequations}
\begin{align}
    [\hat{\beta}_l, \hat{\beta}_m]_\theta &\equiv \hat{\beta}_l \hat{\beta}_m - e^{i\theta\,\text{sgn}(l-m)} \hat{\beta}_m \hat{\beta}_l = 0, \\
    [\hat{\beta}_l, \hat{\beta}_m^\dagger]_{-\theta} &\equiv \hat{\beta}_l \hat{\beta}_m^\dagger - e^{-i\theta\,\text{sgn}(l-m)} \hat{\beta}_m^\dagger \hat{\beta}_l = \delta_{lm}, \label{eq:anyonic_comm}
\end{align}
\end{subequations}
where $\theta$ is the anyonic statistical phase.
$\theta=0$ and $\pi$ represent bosons and ``pseudofermions" (fermions that can occupy the same lattice site as bosons), respectively.
Particles on the same chain obey the anyonic commutation relations with statistical angle $\theta$, whereas particles on different chains commute and carry no mutual statistical phase. This convention is implemented through chain-resolved Jordan–Wigner strings, as detailed in the Supplemental Material~\cite{SuppMat}.

In Fig.~\ref{fig:2}, we show two representative results for $N=2$, corresponding to $\theta=0$ and $\theta=\pi/2$.
Under a strong $U$, two-particle bound states form two energy clusters centered at ${\rm Re}(E)\approx U$, 
with  ${\rm Im}(E)\approx 0$ and $-2\gamma$ separated by the dissipation.
For $\theta=0$, these two clusters possess curve-like spectrum, with spatially uniform particle distribution under OBCs [Figs.~\ref{fig:2}(b) and \ref{fig:2}(c)].
In contrast, a point gap is seen to open under PBCs when $\theta=\pi/2$, reflecting a nontrivial spectral winding topology; and the OBC eigenstates become localized near boundaries, indicating NHSE of the bound states.

As a signature of the point-gap topology, the NHSE can be characterized by a directional inverse participation ratio (dIPR)
~\cite{PhysRevB.106.085427,PhysRevB.105.245407}, defined as
\begin{align}
{\rm dIPR}_m & =\frac{1}{N^2}\sum_l \frac{\left(l-(L+1)/2\right)\sum_{\beta}(\rho_{l,m}^{\beta})^2}{(L-1)/2}, \label{eq:dIPR}
\end{align}
with $N$ the particle number and $\rho_{l,m}^{\beta}=\langle \psi_m| \hat{\beta}^{\dagger}_l\hat{\beta}_l |\psi_m \rangle$, and $|\psi_m\rangle$ the $m$th right eigenstate of the system.
The sign and magnitude of ${\rm dIPR}_m$ characterize the localization direction and strength of $|\psi_m\rangle$, respectively. 
As shown in Figs.~\ref{fig:2}(b) to~\ref{fig:2}(d), 
the continuum states near 
${\rm Re}(E)\approx0$ remain extended (${\rm dIPR}_m\approx0$), while the bound-state clusters 
develop clear directional localization toward opposite directions for $\theta=\pi/2$.

\begin{figure} 
\centering
\includegraphics[width=\linewidth]{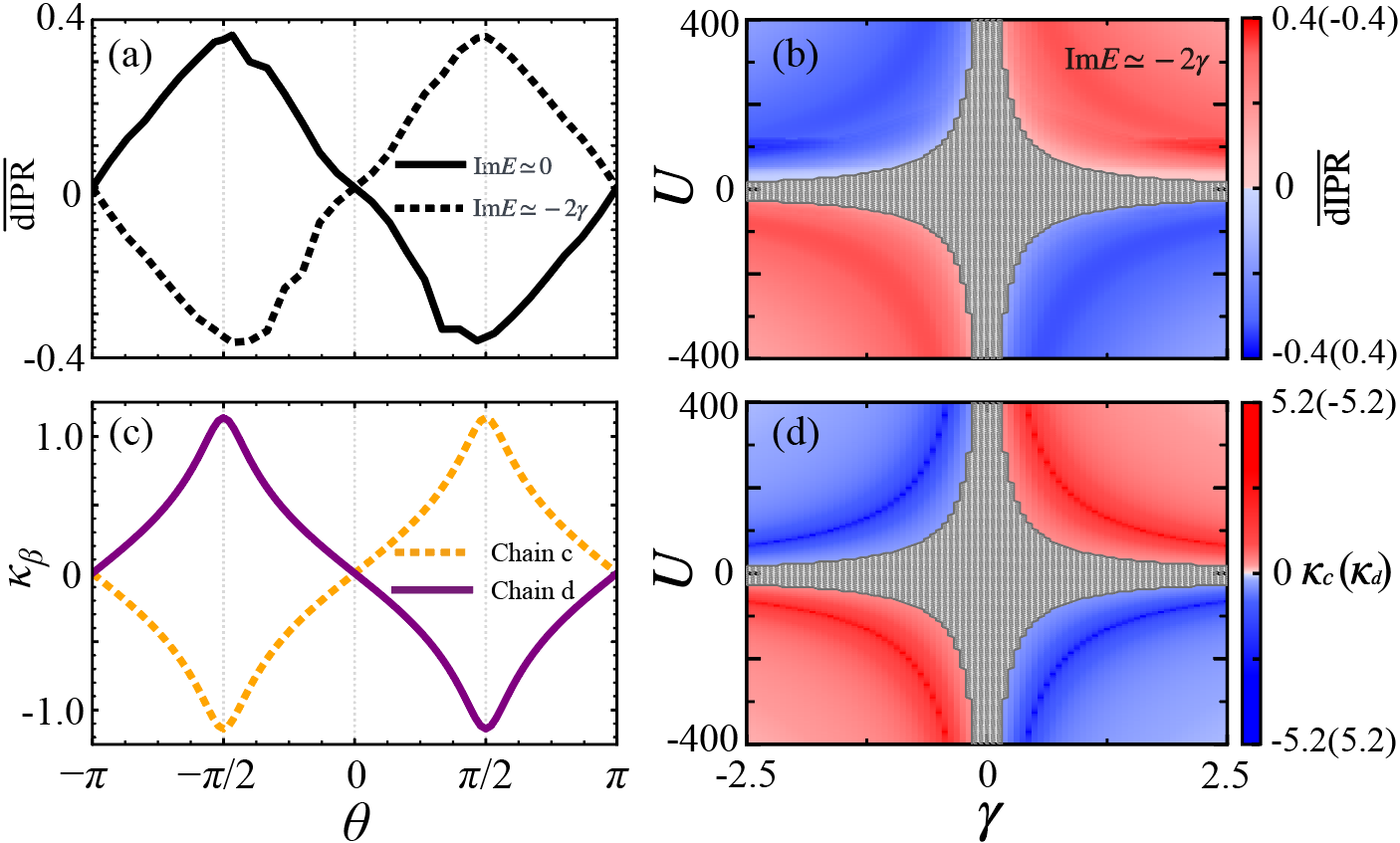}  
\caption{
(a) $\overline{\mathrm{dIPR}}$ of the upper- and lower-imaginary-energy bound-state clusters as a function of the statistical angle $\theta$, with $U=100$ and $\gamma=2$.
The two clusters exhibit opposite NHSE, indicated by nonzero  $\overline{\mathrm{dIPR}}$ with opposite signs when $\theta\not\in\{0,\pi\}$.
(b) Phase diagram of $\overline{\mathrm{dIPR}}$ for the lower cluster in the $(\gamma,U)$ plane with $\theta=\pi/2$.
(c) Effective non-reciprocity $\kappa_{\beta}$ ($\beta=c$ or $d$) [see Eq.~\eqref{eq:kappa}] as a function of $\theta$ for the two transformed chains with the same parameters as for (a),
which shows the same trend as the $\overline{\mathrm{dIPR}}$.
(d) Phase diagram of $\kappa_{c,d}$ in the $(\gamma,U)$ plane with $\theta=\pi/2$.
In (b) and (d), 
the central gray area corresponds to energy-mixture between bound states and scattering states (the two bound-state clusters) at small $U$ ($\gamma$), where $\overline{\rm dIPR}$ becomes ill-defined and $\kappa_{c,d}$ cannot capture the localization properties of the system. Other parameters are the same as in Fig. \ref{fig:2}.
}
\label{fig:3}
\end{figure}

The collective behavior of each cluster is further characterized by the averaged dIPR,
\begin{align}
\overline{\mathrm{dIPR}} &= \frac{1}{M}\sum_{m \in {\rm cluster}} {\rm dIPR}_m \label{eq:dIPR_cluster},
\end{align}
where  $M$ is the total number of eigenstates within a given cluster.
It is seen in Fig.~\ref{fig:3}(a) that $\overline{\mathrm{dIPR}}$ vanishes at the bosonic and pseudofermionic limits ($\theta=0,\pi$) and becomes nonzero otherwise,
verifying the anyon-induced NHSE for the bound states.
A phase diagram regarding $\overline{\rm dIPR}$ is shown in Fig.~\ref{fig:3}(b), where the localization direction is seen to be determined by the signs of the interaction $U$ and the non-Hermitian potential $\gamma$.
In addition, weak dissipation mixes the two bound-state clusters, while weak interaction leaves them admixed with the continuum; either renders $\overline{\mathrm{dIPR}}$ ill-defined. These mixed regimes at small $U$ or $\gamma$ are discussed in the Supplemental Materials~\cite{SuppMat}.

\emph{Origin of the anyon-induced non-Hermitian point gap.---}
To unveil the origin of the anyon-induced point-gap topology, 
we first map the statistical angle $\theta$ to a density-dependent phase factor on hopping terms,
through a generalized Jordan-Wigner transformation~\cite{SuppMat}
\begin{align}
\hat{a}_l = \hat{c}_l e^{- i \theta \sum_{k=1}^{l-1}\hat{a}^\dagger_{k}\hat{a}_{k}}, \quad
\hat{b}_l = \hat{d}_l e^{- i \theta \sum_{k=1}^{l-1}\hat{b}^\dagger_{k}\hat{b}_{k}}, \label{eq:JW}
\end{align}
with $\hat{c}_l$ and $\hat{d}_l$ satisfying bosonic commutation relations.
The unitarity ensures the same particle distribution after the transformation.
Next, since the point-gap topology occurs only for the two bound-state clusters, 
we project the Hamiltonian onto the basis of $N$-particle bound states,
$\ket{\alpha_{\beta,l}}=
\frac{1}{\sqrt{N!}}\left(\hat\beta_l^\dagger\right)^N\ket 0$, with $|0\rangle$ the vacuum state and $\alpha\in\{c,d\}$. 
Treating the hopping terms as perturbations,
the leading nonvanishing processes arise at $N$th order, and an effective single-particle Hamiltonian can be obtained as (omitting a uniform energy shift)~\cite{SuppMat}
\begin{widetext}
\begin{align}
\hat H_{\rm proj}^{(N)}
=&
\sum_{l=1}^{L}
\Bigg[\sum_{\beta=c,d}
J_{\Phi}^{(N)}\hat\alpha_{\beta,l}^{\dagger}\hat\alpha_{\beta,l+1}+
J_0^{(N)}\hat\alpha_{c,l}^{\dagger}\hat\alpha_{d,l}
+
J_1^{(N)}
\Big(
\hat\alpha_{c,l}^{\dagger}\hat\alpha_{d,l+1}
+
\hat\alpha_{d,l}^{\dagger}\hat\alpha_{c,l+1}
\Big)
\Bigg]+\mathrm{h.c.} 
+\sum_{l=1}^{L}\sum_{\beta=c,d}
U_\beta^{(N)}\hat\alpha_{\beta,l}^{\dagger}\hat\alpha_{\beta,l},
\label{eq:H_eff1}
\end{align}
\end{widetext}
with $\{J_{\Phi}^{(N)}, U_c^{(N)},U_d^{(N)}\}\in\mathbb{C}$ and $\{J_0^{(N)},J_1^{(N)}\}\in\mathbb{R}$.
Here the perturbed hopping parameters are proportional to the $N$th power of the original ones, and the onsite potentials $U_{c,d}^{(N)}$ encode the non-Hermiticity (see Supplemental Materials~\cite{SuppMat} for their explicit forms).
In particular, 
the fractional statistics manifests only in
\begin{align}
J_{\Phi}^{(N)}
=
J^{(N)}e^{-i\Phi_N},~~J^{(N)}=(-1)^N\frac{N!}{[(N-1)!]^2}
\frac{J^N}{U^{N-1}},\label{eq:J_N}
\end{align}
with $\Phi_N=\frac{N(N-1)}{2}\theta$ 
the total statistical phase accumulated when moving an $N$-particle bound state by one site, during which each particle exchanges with the others via the density-dependent phase in Eq.~\eqref{eq:JW}.

With this single-particle picture, the point-gap topology can be revealed through the Bloch form of $\hat H_{\rm proj}^{(N)}$,
\begin{align}
h(k)=h_0(k)\hat{\sigma}_0+\sum_{s=x,y,z}h_s(k)\hat{\sigma}_s ,
\label{eq:hk}
\end{align}
where
$h_0(k)=(U_c^{(N)}+U_d^{(N)})/2+2J^{(N)}\cos(k-\Phi_N)$,
$h_x(k)=J_0^{(N)}+2J_1^{(N)}\cos k$, $h_y(k)=0$, and
$h_z(k)=(U_c^{(N)}-U_d^{(N)})/2$. 
The eigenenergies are
\begin{align}
E_\pm(k)=h_0(k)\pm \Delta E(k),~~
\Delta E(k)=\sqrt{h_x^2(k)+h_z^2},
\label{eq:E_hk}
\end{align}
which acquire $k$-dependence in their real parts via $h_0(k)$, and in both imaginary and real parts via $\Delta E(k)$ ($h_z$ is purely imaginary; see Eqs.~S44 to S46 in Supplemental Materials~\cite{SuppMat}).
Via the statistical phase $\Phi_N$, a nontrivial point-gap is opened by the misalignment between  $h_0(k)$ and $\Delta E(k)$, which play the  roles of real and imaginary energies in Fig. \ref{fig:1}.
The point-gap topology can be trivialized under certain conditions. 
Finally, the opposite signs in $\Delta E(k)$ of the two bands further indicate their opposite spectral winding numbers and NHSE directions,
as shown in Fig.~\ref{fig:2}(e).

Conventionally, point-gap topology can be characterized by a spectral winding number defined for a chosen reference energy enclosed by the spectral loop~\cite{PhysRevX.9.041015,PhysRevLett.124.086801,PhysRevLett.125.126402}.
This makes the topological invariant not uniquely defined in our model, where loop spectra drift with parameters on the complex energy plane. Instead, we use the signed spectral area of each band for characterizing the spectral winding topology, defined as
\begin{align}
S_\pm
=\,
\mp 2J^{(N)}
&\sin\Phi_N
\int_{-\pi}^{\pi} v_N(k)\cos k\,dk,
\nonumber\\
v_N(k)
=&{\rm Im}\sqrt{h_x^2(k)+h_z^2}=\,{\rm Im}[\Delta E(k)],
\label{eq:area_main}
\end{align}
for the two bands of $E_\pm(k)$~\cite{SuppMat}.
$S_{\pm}$ provides a topological invariant for each band through its sign: 
${\rm Sgn}[S_{\pm}]=\pm 1$ indicates the domination of positive and negative spectral winding number for the corresponding energy band, respectively; and $S_{\pm}=0\Rightarrow {\rm Sgn}[S_{\pm}]=0$ indicates vanishing spectral topology, resulting from either the point-gap closing, or zero net-winding for a single band from the cancellation between  positive and negative winding areas with the same weight.
The parameter conditions for $S_\pm=0$, and their corresponding physical interpretations, are summarized in Table \ref{tab:area_zero_conditions} (see Supplemental Materials~\cite{SuppMat} for more details).



\begin{table}[t]
\caption{Parameter conditions leading to $S_\pm=0$ (vanishing point-gap or cancellation between oppositely oriented lobes) and their interpretations.}
\label{tab:area_zero_conditions}
\begin{tabular*}{\columnwidth}{@{\extracolsep{\fill}}cc}
\hline\hline
\textbf{Parameter condition} & \textbf{Interpretation} \\
\hline
$\Phi_N = 0\,({\rm mod}~\pi)$ 
& trivial statistical angle$^*$ \\
$U\rightarrow\infty$ & Atomic limit \\
$\gamma=0$ \;($\mathrm{Im}(h_z)=0$) & Hermitian limit \\
$J=0$ or $J_1=0$ & $\Phi_N$ deactivated$^\dagger$ \\
$J_0=0$ & Symmetric spectral trajectory$^{\ddagger}$ \\
\hline\hline
\end{tabular*} 
\begin{tablenotes}     
\item $*$ This condition corresponds to bosons and pseudofermions ($\theta=0$ and $\pi$) when $N=2$, and arbitrary $\theta$ for $N=1$.
\item $^\dagger$  $\Phi_N$ induces a point-gap topology through the mismatch between these two terms; see Eq.~\eqref{eq:hk} and following discussion. 
\item $\ddagger$ This does not preclude nontrivial spectral winding, but yields a figure-eight spectrum possessing equivalent positive and negative winding areas that cancel each other in a single band~\cite{SuppMat}.
    \end{tablenotes}
\end{table}

\begin{figure} 
\centering
\includegraphics[width=\linewidth]{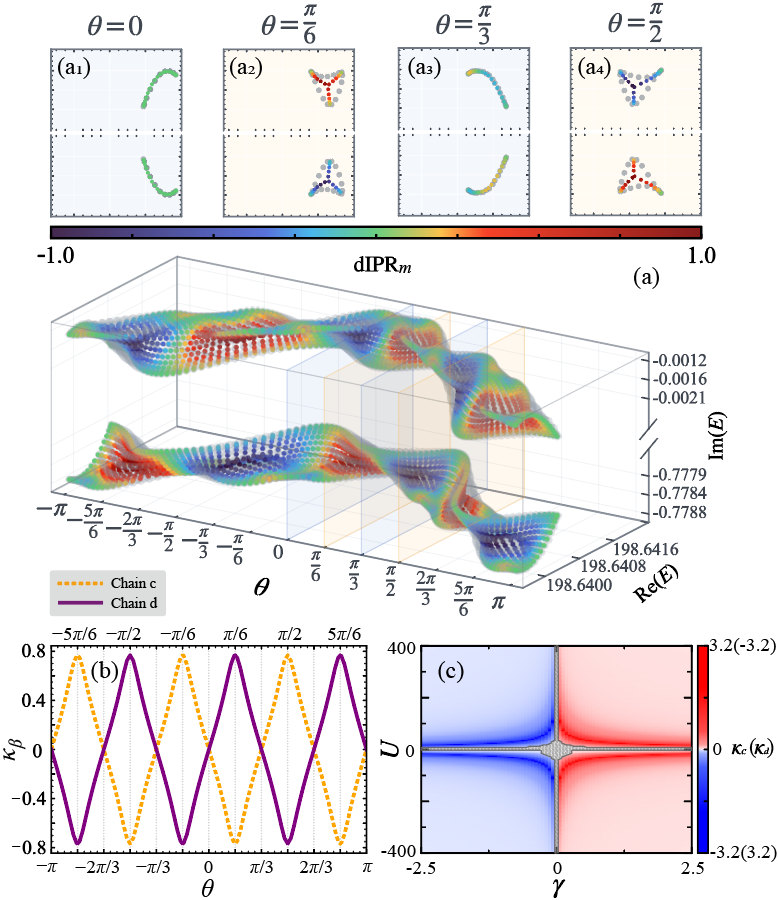}  
\caption{
Anyon-induced non-Hermitian topology for $N=3$,
(a) Complex spectra of the three-particle bound-state clusters at Re$(E)\approx 3U$, versus the statistical angle $\theta$. OBC eigenstates are colored by ${\rm dIPR}_m$; representative cuts are shown for $\theta=0,\pi/6,\pi/3,\pi/2$.
Nontrivial point-gap loops appear except when $\Phi_3=3\theta=m\pi$.
(b) Effective non-reciprocity $\kappa_{c,d}^{(3)}$ as a function of $\theta$, showing the accelerated oscillation governed by $\Phi_3=3\theta$ and the opposite signs of the two effective chains.
(c) Phase diagram of $\kappa_{c,d}^{(3)}$ in the $(\gamma,U)$ plane at $\theta=\pi/6$.
Data at small $U$ ($\gamma$) are omitted as eigenenergies are mixed between bound states and scattering states (the two bound-state clusters).
For (a) and (b), $U=66$ and $\gamma=0.26$.
Other parameters are $L=20$, $N=3$, $J_0=J_1=3$, and $J=1$.
}
\label{fig:4}
\end{figure}

\emph{$\theta$-dependence and particle-number parity effect for larger $N$.---}
The particle number $N$ crucially determines the cumulative statistical phase $\Phi_N$ and other $N$th-order parameters in Eq.~\eqref{eq:H_eff1}, which together govern the point-gap topology.
This can already be seen from the signed area $S_\pm$ in Eq.~\eqref{eq:area_main}.
To quantitatively describe these effects, 
we extract the non-reciprocity for each cluster by further constructing effective single-chain Hamiltonians, treating  inter-chain couplings $J_0^{(N)}$ and $J_1^{(N)}$ perturbatively.
This approach is justified as eigenstates in each cluster predominantly reside on one chain due to the separation in ${\rm Im}(E)$ [e.g., see Fig. \ref{fig:2}]. 
Explicitly, the single-chain non-reciprocity can be characterized by an effective inverse localization length of the skin modes~\cite{SuppMat},
\begin{align}
\kappa_{c}^{(N)} = \frac{1}{4}\ln\!\left[\frac{(J^{(N)})^2 + Y^2 - 2  J^{(N)}Y \sin\Phi_N}{(J^{(N)})^2 + Y^2 + 2  J^{(N)} Y \sin\Phi_N}
\right] = -\kappa_{d}^{(N)}, \label{eq:kappa}
\end{align}
with $Y=2J_0^{(N)}J_1^{(N)}/Z$, and 
$Z=-N\gamma+2\gamma\left(NJ_0^2+2NJ_1^2\right)/\left((N-1)^2U^2+\gamma^2\right).$
As can be seen in Figs.~\ref{fig:3}(c) and \ref{fig:3}(d), $\kappa_{c,d}^{(N)}$ show the same trend as dIPR and capture the localization signatures.

The sign of $\kappa_{c,d}^{(N)}$ indicates the non-reciprocal direction, which corresponds to the sign of the spectral winding number characterizing the point-gap~\cite{PhysRevLett.124.086801,PhysRevLett.125.126402}.
Non-trivial point-gap topology arises when $\kappa_{c,d}^{(N)}\neq0$, which requires $\sin\Phi_N\neq0$ in Eq.~\eqref{eq:kappa}.
Thus, a larger $N$ leads to more delicate dependence of the point-gap topology on the statistical phase $\theta$.
Figure~\ref{fig:4} demonstrates an example with $N=3$.
Specifically, the two clusters of three-particle bound states at Re$(E)\approx 3U$ are seen to possess nontrivial point-gap topology and opposite NHSE for general $\theta$, except for $\Phi_3=3\theta=0$ or $\pi$, as shown in Fig. \ref{fig:4}(a).
Consistently, $\kappa_{c,d}^{(3)}$ in Fig.~\ref{fig:4}(b) oscillate with $\theta$ and become zero when the point-gap topology vanishes.

Finally, in Fig. \ref{fig:4}(c), we demonstrate the phase diagram in the $(\gamma,U)$ space for $N=3$. 
$\kappa_{c,d}^{(N)}$ does not change its sign with $U$,
unlike the case for $N=2$ in Fig. \ref{fig:3}(d).
Such $N$-parity dependence of the point-gap topology arises from
the $1/U^{N-1}$ factor of $J^{(N)}$ in Eq.~\eqref{eq:J_N}, so that the sign of $U$ determines the sign of $\kappa_{c,d}^{(N)}$ in Eq.~\eqref{eq:kappa}, altering the non-reciprocity only under even $N$.
We also note that, due to the $1/U^{N-1}$ factor,
the point-gap topology becomes weaker as $N$ increases, as reflected by the narrow energy range of the bound states in Fig.~\ref{fig:4}(a). Nevertheless, the inverse localization length $\kappa_{\beta}$ reaches comparable maximum values for $N=2$ ($\sim 5.2$) and $N=3$ ($\sim 3.2$), indicating similarly strong skin localization in appropriate parameter regimes.

\emph{Conclusion and discussion.---}
With a 1D ladder example, we have shown that anyonic statistics can induce non-Hermitian point-gap topology and corresponding bound-state NHSE in the presence of sublattice-dependent dissipation.
The conditions for realizing nontrivial point-gap topology are identified through analyzing the signed spectral area, whose sign characterizes the topology of each band as a whole.
Notably, the anyon-induced point-gap topology and bound-state NHSE are intimately tied to the particle number $N$, with the skin localization oscillating rapidly with the statistical phase and exhibiting a parity dependence on $N$.
These results reveal fractional statistics as an intrinsic origin of non-Hermitian topological phases,
opening a route toward engineering novel topological phases and the corresponding boundary phenomena from quantum statistics.
Experimentally, ultracold atoms provide a promising platform for realizing our model and the predicted phenomena that rely on quantum many-body effects. In particular, 1D anyonic behavior has recently been realized through Floquet-engineered density-dependent phases~\cite{kwan2024realization} and spin–charge separation~\cite{dhar2025observing}, while NHSE associated with point-gap topology has also been achieved via near-resonant optical transition~\cite{liang2022dynamic,zhao2025two}.
Our numerical simulations also show clear non-reciprocal dynamics corresponding to the anyon-induced point-gap topology (see End Matter).

Notably, while here we focus on a specific 1D model, the underlying mechanism may be extended to other models containing sublattice/pseudospin-dependent gain and loss.
This is because, through the generalized Jordan-Wigner transformation, the anyonic statistical phase acts as a gauge flux that can induce a chiral current to the system. Under pseudospin-dependent gain and loss, the chiral current toward opposite direction suffers different suppression or enhancement, leading to effective non-reciprocity that can induce point-gap topology and NHSE~\cite{PhysRevLett.128.223903}.
Thus, a promising direction is the extension to other classes of systems, especially higher dimensions, potentially enabling non-Abelian anyonic statistics and nontrivial braiding structures.


\begin{center}
\textbf{Acknowledgement}
\end{center}
L. Li acknowledges helpful discussion with Y. Qin and C. H. Lee.
This work is supported by
the National Natural Science Foundation of China (Grants No. 12474159 and No. 12304315), 
the National Key R\&D Program of China (No. 2022YFA1404500, No. 2022YFA1404701),  
the Guangdong Provincial Quantum Science Strategic Initiative (Grants No. GDZX2504003 and GDZX2504006),
and the Shanghai Science and Technology Innovation Action Plan (No. 24Z510205936).

\clearpage


\begin{center}
\textbf{End Matter}
\end{center}

\begin{figure}[t]
\centering
\includegraphics[width=\linewidth]{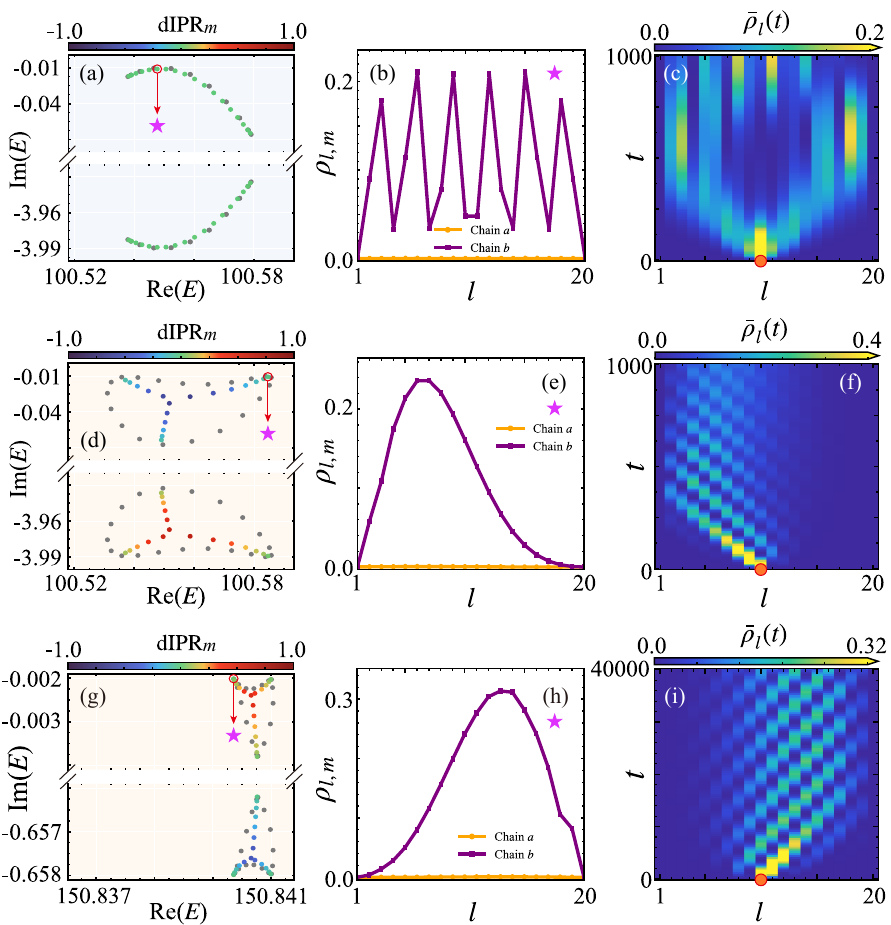}
\caption{
Dynamics of the bound-state.
(a) to (c) Bosonic limit, $\theta=0$, with $N=2$, $U=100$, and $\gamma=2$: 
(a) Complex bound-state spectra, with PBC eigenvalues shown as gray points, OBC
eigenvalues colored by \({\rm dIPR}_m\), and the magenta star marking the
eigenstate with the largest \({\rm Im}(E)\); 
(b) density distribution $\rho_{l,m}=\langle\psi_m|\hat n_{\beta,l}|\psi_m\rangle$ of the marked state; 
(c) evolution of $\bar{\rho}_{l}(t)$ from $\ket{\psi(0)}=\ket{b_{10},b_{10}}$. 
Orange dot marks the initial state.
(d) to (f) Anyonic case at $\theta=\pi/2$, with the same parameters and initial state. 
(g) to (i) $N=3$ anyonic case at $\theta=\pi/6$, $U=50$, and $\gamma=0.22$, with $\ket{\psi(0)}=\ket{b_{10},b_{10},b_{10}}$.
Other parameters are \(L=20\), \(J=1\), and \(J_0=J_1=3\).
}
\label{fig:EM_dynamics}
\end{figure}

\emph{Dynamics of the bound-state.---}
The phenomena discussed in the main text can also be captured from the dynamical properties of the system. We consider the time evolution of an initial state \(\ket{\psi(0)}\), in which all particles are located at the central site. At time \(t\), the normalized evolved state is given by
\begin{align}
\ket{\psi(t)}
=
\frac{e^{-i\hat H_{\rm a}t}\ket{\psi(0)}}
{\sqrt{\bra{\psi(0)}e^{i\hat H_{\rm a}^{\dagger}t}
e^{-i\hat H_{\rm a}t}\ket{\psi(0)}}}.
\end{align}
We define the normalized averaged density distribution over the two chains as $
\bar{\rho}_{l}(t)
=
1/N \sum_{\beta=a,b}
\bra{\psi(t)}\hat n_{\beta,l}\ket{\psi(t)} .$
In the bosonic limit \(\theta=0\), the bound-state OBC spectrum has \({\rm dIPR}_{m}\approx 0\), and the PBC spectrum does not form spectral loops [Fig.~\ref{fig:EM_dynamics}(a)]. Starting from the initial state
\(\ket{\psi(0)}=(\hat b_{10}^{\dagger})^{2}\ket{0}/\sqrt{2}\),
the density distribution disperses in space over time [Fig.~\ref{fig:EM_dynamics}(c)]. At sufficiently long times, the density profile approaches that of the eigenstate with the largest imaginary part in the full spectrum [Fig.~\ref{fig:EM_dynamics}(b)].

By contrast, in the anyonic case, for instance at \(\theta=\pi/2\), the two clusters in the bound-state sector form oppositely oriented spectral loops. Consequently, the initial state
\(\ket{\psi(0)}=(\hat b_{10}^{\dagger})^{2}\ket{0}/\sqrt{2}\)
first drifts to the left at short times, following the NHSE direction of the cluster with larger imaginary energies. In the long-time limit, the density distribution again converges to that of the eigenstate with the largest imaginary part in the full spectrum [Figs.~\ref{fig:EM_dynamics}(d)--\ref{fig:EM_dynamics}(f)].

For the $N=3$ anyonic case, we choose \(\theta=\pi/6\), corresponding to the first peak of $\kappa_{\beta}^{(N)}$ ($\beta=c$ or $d$) as \(\theta\) is increased from zero [see Fig.~\ref{fig:4}(b)], in the same manner as for \(N=2\). The two bound-state clusters wind in directions opposite to those of the corresponding clusters for \(N=2\). As a result, the initial state
\(\ket{\psi(0)}=(\hat b_{10}^{\dagger})^{3}\ket{0}/\sqrt{3!}\)
first drifts to the right at short times, again following the NHSE direction of the cluster with larger imaginary energies. 
At long times, the density distribution approaches that of the eigenstate with the largest imaginary part in the full spectrum [Figs.~\ref{fig:EM_dynamics}(g)--\ref{fig:EM_dynamics}(i)].
We also note that the evolution is much slower compared to the case with $N=2$, as the bound-state NHSE corresponds to higher-order hopping processes for larger $N$, where the effective hopping becomes weaker [see Eq.~\eqref{eq:J_N} for an example].

\clearpage
\onecolumngrid    
\appendix

\newpage
\begin{center}
\textbf{\large Supplemental Materials}
\end{center}

\tableofcontents
\setcounter{secnumdepth}{2}
\setcounter{equation}{0} \setcounter{figure}{0} 
\setcounter{table}{0} %
\renewcommand{\theequation}{S\arabic{equation}} 
\renewcommand{\thefigure}{S%
\arabic{figure}} 

\phantomsection
\subsection*{I. Effective single-particle Hamiltonian for bound states}
In this section, we first map the anyonic system to a bosonic one with density-dependent phase factors, then obtain the effective single-particle Hamiltonian (Eq.~(\ref{eq:H_eff1}) of the main text) by projecting the Hamiltonian to the sub-Hilbert space of $N$-particle bound states.
Finally, we also briefly discuss the Tamm-Shockley boundary states originating from interaction-induced effective boundary impurities in our system.
\subsubsection*{A. Jordan--Wigner transformation and bosonic representation}\label{s:sec:JW_boson}
The one-dimensional anyonic Creutz ladder model we consider is described by the Hamiltonian
\begin{align}
\hat{H}_{\rm{a}} = 
& - \sum_{l=1}^L \Bigg[
    \sum_{\alpha=a,b} J \hat{\alpha}_l^\dagger \hat{\alpha}_{l+1} 
    + J_1 \left( \hat{a}_l^\dagger \hat{b}_{l+1} + \hat{b}_l^\dagger \hat{a}_{l+1} \right) 
    + J_0 \hat{a}_l^\dagger \hat{b}_l 
\Bigg] \notag\\
&+ \text{h.c.}
+ \frac{U}{2} \sum_{\alpha=a,b} \sum_{l=1}^L 
    \hat{n}_{\alpha,l} \left( \hat{n}_{\alpha,l} - 1 \right) 
- i \gamma \sum_{l=1}^L \hat{a}_l^\dagger \hat{a}_l,
 \label{s:eq:anyon_H}
\end{align}
with
\begin{subequations}
\begin{align}
    &\hat{\beta}_l \hat{\beta}_m - e^{i\theta\,\text{sgn}(l-m)} \hat{\beta}_m \hat{\beta}_l = 0, \qquad
   \hat{\beta}_l \hat{\beta}_m^\dagger - e^{-i\theta\,\text{sgn}(l-m)} \hat{\beta}_m^\dagger \hat{\beta}_l = \delta_{lm},\\
&[\hat{\beta}_l,\hat{\bar{\beta}}_m]=[\hat{\beta}_l^\dagger,\hat{\bar{\beta}}_m]=[\hat{\beta}_l^\dagger,\hat{\bar{\beta}}_m^\dagger]=0,
\end{align}
\end{subequations}
where $\theta$ is the anyonic statistical phase, $\beta,\bar{\beta}\in\{a,b\}$, and $\beta\neq\bar{\beta}$.
We then consider a generalized Jordan--Wigner transformation
\begin{align}
\hat{a}_l =
\hat{c}_l e^{- i \theta \sum_{k=1}^{l-1}\hat{n}_{c,k}},
\qquad
\hat{b}_l =
\hat{d}_l e^{- i \theta \sum_{k=1}^{l-1}\hat{n}_{d,k}},
\end{align}
where the transformed particle creation and annihilation operators ($\hat{c}_l$, $\hat{d}_l$, $\hat{c}^\dagger_l$, and $\hat{d}^\dagger_l$) satisfy bosonic commutation relations.
Applying the transformation, we obtain
\begin{align}\label{s:eq:JW1}
	-\sum_{l=1}^{L}\sum_{\alpha = a,b} J \hat{\alpha}_l^{\dagger}\hat{\alpha}_{l+1} &= - \sum_{l=1}^{L}\sum_{\beta = c,d}J e^{i\theta \sum_{k=1}^{l-1}\hat{n}_{\beta,k}} \hat{\beta}_{l}^{\dagger} \hat{\beta}_{l+1}e^{-i\theta \sum_{k=1}^{l}\hat{n}_{\beta,k}}\cr
	&= - \sum_{l=1}^{L}\sum_{\beta = c,d}J \hat{\beta}_{l}^{\dagger} e^{i\theta \sum_{k=1}^{l-1}\hat{n}_{\beta,k}} e^{-i\theta \sum_{k=1}^{l}\hat{n}_{\beta,k}}\hat{\beta}_{l+1} \cr
    &= - \sum_{l=1}^{L}\sum_{\beta = c,d}J \hat{\beta}_{l}^{\dagger} e^{-i\theta \hat{n}_{\beta,l}}\hat{\beta}_{l+1}
\end{align}

\begin{align}\label{s:eq:JW2}
	-\sum_{l=1}^{L}J_1 \big( \hat{a}_l^\dagger \hat{b}_{l+1} + \hat{b}_l^\dagger \hat{a}_{l+1} \big)  &= -\sum_{l=1}^{L} J_1 \big(  \hat{c}_l^{\dagger}e^{i\theta \sum_{k=1}^{l-1}(\hat{n}_{c,k} - \hat{n}_{d,k} ) - i\theta \hat{n}_{d,l}} \hat{d}_{l+1} \cr
    &+ \hat{d}_l^{\dagger}e^{i\theta \sum_{k=1}^{l-1}(\hat{n}_{d,k} - \hat{n}_{c,k} ) - i\theta \hat{n}_{c,l}} \hat{c}_{l+1} \big) \cr
	&=-\sum_{l=1}^{L} J_1 \big(  \hat{c}_l^{\dagger}e^{i\theta (\Delta_{cd}^{(l-1)}-  \hat{n}_{d,l})} \hat{d}_{l+1}\cr
    &+ \hat{d}_l^{\dagger}e^{-i\theta(\Delta_{cd}^{(l-1)} + \hat{n}_{c,l})} \hat{c}_{l+1} \big)
\end{align}

\begin{align}\label{s:eq:JW3}
	-\sum_{l=1}^{L} J_{0} \hat{a}_l^{\dagger} \hat{b}_{l} &= - \sum_{l=1}^{L} J_0  \hat{c}_{l}^{\dagger} e^{i\theta \sum_{k=1}^{l-1}\hat{n}_{c,k}} e^{-i\theta \sum_{k=1}^{l-1}\hat{n}_{d,k}} \hat{d}_{l} \cr
	&=- \sum_{l=1}^{L} J_0  \hat{c}_{l}^{\dagger} e^{i\theta \Delta_{cd}^{(l-1)}}  \hat{d}_{l} 
\end{align}

\begin{align}\label{s:eq:JW4}
	 \frac{U}{2} \sum_{\alpha=a,b} \sum_{l=1}^L \hat{n}_{\alpha,l} \big( \hat{n}_{\alpha,l} - 1 \big) = \frac{U}{2} \sum_{\beta = c,d} \sum_{l=1}^L \hat{n}_{\beta,l}(\hat{n}_{\beta,l} - 1)
\end{align}

\begin{align}\label{s:eq:JW5}
	- i \gamma \sum_{l=1}^L \hat{a}_l^\dagger \hat{a}_l = - i \gamma \sum_{l =1}^{L} \hat{c}_l^\dagger \hat{c}_l
\end{align}
Substituting Eqs.~\eqref{s:eq:JW1}--\eqref{s:eq:JW5} into Eq.~\eqref{s:eq:anyon_H}, we obtain a bosonic system with density-dependent phase factors,
\begin{align}\label{s:eq:Boson_H}
\hat{H}_{\rm{b}}
=&-\Bigg\{
\sum_{l=1}^{L-1}\left[
\sum_{\beta = c,d}J \hat{\beta}_{l}^{\dagger}e^{-i \theta \hat{n}_{\beta,l}}\hat{\beta}_{l+1}
+J_1 \hat{c}^{\dagger}_l e^{i \theta(\Delta_{cd}^{(l-1)} - \hat{n}_{d,l})} \hat{d}_{l+1}
+J_1 \hat{d}^{\dagger}_{l} e^{-i \theta(\Delta_{cd}^{(l-1)} + \hat{n}_{c,l})} \hat{c}_{l+1}
\right]\nonumber\\
&\qquad\qquad
+\sum_{l=1}^{L}J_0 \hat{c}^{\dagger}_{l} e^{i \theta \Delta_{cd}^{(l-1)}} \hat{d}_{l}
\Bigg\}+\mathrm{h.c.}\nonumber\\
&+\sum_{l=1}^{L}\sum_{\beta = c,d}\frac{U}{2} \hat{n}_{\beta,l}(\hat{n}_{\beta,l} - 1) - i \gamma \sum_{l =1}^{L} \hat{n}_{c,l},
\end{align}
where $\Delta_{cd}^{(l-1)}=\sum_{j=1}^{l-1}\left(\hat n_{c,j}-\hat n_{d,j}\right)$.

\phantomsection
\subsubsection*{B. Projected Hamiltonian for \texorpdfstring{\(N\)-particle}{N-particle} bound states}
\label{s:sec:N_particle_bound_state_projection}
Next we project Eq.~\eqref{s:eq:Boson_H} onto the $N$-particle bound-state subspace.
We treat the hopping part as perturbation,
\begin{align}
    \hat{H}_{\rm hop}
    =&-\Bigg\{
    \sum_{l=1}^{L-1}
    \left[
    \sum_{\beta = c,d}J
    \hat{\beta}_{l}^{\dagger}e^{-i \theta \hat{n}_{\beta,l}}\hat{\beta}_{l+1}
    +J_1
    \hat{c}^{\dagger}_{l} e^{i \theta(\Delta_{cd}^{(l-1)} - \hat{n}_{d,l})} \hat{d}_{l+1}
    +J_1
    \hat{d}^{\dagger}_{l} e^{-i \theta(\Delta_{cd}^{(l-1)} + \hat{n}_{c,l})} \hat{c}_{l+1}
    \right]\nonumber\\
    &\qquad\qquad
    +\sum_{l=1}^{L}J_0
    \hat{c}^{\dagger}_{l} e^{i \theta \Delta_{cd}^{(l-1)}} \hat{d}_{l}
    \Bigg\}+\mathrm{h.c.},
    \label{s:eq:Boson_H_hop}
\end{align}
and the onsite interaction and loss part as the unperturbed Hamiltonian,
\begin{align}
    \hat{H}_{\rm int}
    =
    \sum_{l=1}^{L}\sum_{\beta = c,d} \frac{U}{2} \hat{n}_{\beta,l}(\hat{n}_{\beta,l} - 1)
    - i \gamma \sum_{l =1}^{L} \hat{n}_{c,l}.
    \label{s:eq:Boson_H_int}
\end{align}
For an $N$-particle bound state, we define
\begin{align}
\ket{\alpha_{\beta,l}}
=
\frac{1}{\sqrt{N!}}\left(\hat\beta_l^\dagger\right)^N\ket 0,
\qquad \beta=c,d.
\label{s:eq:N_bound_basis}
\end{align}
The corresponding intermediate states along the ordered virtual path are
\begin{align}
\ket{m^{\beta}_{k,l}}
=
\frac{1}{\sqrt{(N-k)!\,k!}}
\left(\hat\beta_l^\dagger\right)^k
\left(\hat\beta_{l+1}^\dagger\right)^{N-k}\ket 0,
\qquad k=0,1,\dots,N,
\label{s:eq:N_chain_states}
\end{align}
for the intra-chain hopping channel ($J$), and
\begin{align}
\ket{m^{c\to d}_{k,l}}
=
\frac{1}{\sqrt{(N-k)!\,k!}}
\left(\hat c_l^\dagger\right)^{N-k}
\left(\hat d_{x}^\dagger\right)^k\ket 0,
\qquad k=0,1,\dots,N,
\label{s:eq:N_flip_states}
\end{align}
for the inter-chain hopping channel ($J_0$ and $J_1$), where $x=l$ for $J_0$ and $x=l+1$ for $J_1$.
The reverse $d\to c$ virtual path is obtained analogously by exchanging $c$ and $d$.

A complete moving of the $N$-particle bound state by one site is an $N$th-order hopping process. 
Based on quasi-degenerate perturbation theory, the nonzero matrix elements of the projected Hamiltonian are given by
\begin{align}
\langle \alpha_{\beta,l}|\hat H_{\rm proj}^{(N)}|\alpha_{\beta',l'}\rangle
=
\frac12
\sum_{m_1,\dots,m_{N-1}}
\Bigg[
\frac{
V_{\alpha_{\beta,l}m_1}V_{m_1m_2}\cdots V_{m_{N-1}\alpha_{\beta',l'}}
}{
\prod_{i=1}^{N-1}(E_{\alpha_{\beta,l}}-E_{m_i})
}
+
\frac{
V_{\alpha_{\beta,l}m_1}V_{m_1m_2}\cdots V_{m_{N-1}\alpha_{\beta',l'}}
}{
\prod_{i=1}^{N-1}(E_{\alpha_{\beta',l'}}-E_{m_i})
}
\Bigg],
\label{s:eq:general_symmetric_PT}
\end{align}
where $V\equiv \hat H_{\rm hop}$. For each nonzero effective hopping amplitude evaluated below, only one ordered sequence of virtual intermediate states contributes. Hence the sums over $m_1,\ldots,m_{N-1}$ in Eq.~\eqref{s:eq:general_symmetric_PT} reduce to a single virtual path.

\paragraph{Intra-chain hopping channel.}

We first derive the intra-chain hopping amplitude entering $J_{\Phi}^{(N)}$. Starting from $\ket{\alpha_{\beta,l+1}}$, the $k$-th intermediate state is $\ket{m^{\beta}_{k,l}}$, and the elementary matrix element is
\begin{align}
\langle m^{\beta}_{k+1,l}|\hat H_{\rm hop}|m^{\beta}_{k,l}\rangle
=
- J\sqrt{(k+1)(N-k)}\,e^{-i k\theta}.
\label{s:eq:elementary_J_left}
\end{align}
The corresponding energy difference is
\begin{align}
E_{\alpha_\beta}-E_{m^{\beta}_{k,l}}
=
k(N-k)U,
\qquad k=1,\dots,N-1.
\label{s:eq:energy_diff_J}
\end{align}
Therefore,
\begin{align}
J_{\Phi}^{(N)}
&=
\frac{
\prod_{k=0}^{N-1}
\langle m^{\beta}_{k+1,l}|\hat H_{\rm hop}|m^{\beta}_{k,l}\rangle
}{
\prod_{k=1}^{N-1}(E_{\alpha_\beta}-E_{m^{\beta}_{k,l}})
}
\nonumber\\
&=
\frac{
(-J)^N
\left[\prod_{k=0}^{N-1}\sqrt{(k+1)(N-k)}\right]
e^{-i\theta\sum_{k=0}^{N-1}k}
}{
\prod_{k=1}^{N-1}k(N-k)U
}
\nonumber\\
&=
(-1)^N \frac{N!}{[(N-1)!]^2}\,
\frac{J^N}{U^{N-1}}\,
e^{-i\frac{N(N-1)}{2}\theta}.
\label{s:eq:Jphi_final}
\end{align}
This is the same intra-chain hopping coefficient used in the main text,
\begin{align}
J_{\Phi}^{(N)}
=
J^{(N)}e^{-i\Phi_N},
\qquad
J^{(N)}
=
(-1)^N\frac{N!}{[(N-1)!]^2}
\frac{J^N}{U^{N-1}},
\qquad
\Phi_N = \frac{N(N-1)}{2}\theta.
\label{s:eq:Jphi_def}
\end{align}
The reversed hopping process is the Hermitian-conjugate counterpart,
\begin{align}
\bigl[J_{\Phi}^{(N)}\bigr]^*
=
J^{(N)}e^{+i\Phi_N}.
\label{s:eq:Jphi_hc}
\end{align}
\paragraph{Inter-chain hopping channel.}
For the inter-chain hopping channel, we define the generic hopping strength $J_x$ with $x=0$ or $1$. The elementary matrix element along the chain
$\ket{m^{c\to d}_{k,l}}\to \ket{m^{c\to d}_{k+1,l}}$ is
\begin{align}
\langle m^{c\to d}_{k+1,l}|\hat H_{\rm hop}|m^{c\to d}_{k,l}\rangle
=
- J_x \sqrt{(k+1)(N-k)}.
\label{s:eq:elementary_Jx}
\end{align}
For the intermediate state with $k$ converted particles, the energies satisfy
\begin{align}
E_{\alpha_c}-E_{m^{c\to d}_{k,l}}
=
k\big[(N-k)U-i\gamma\big],
\qquad
E_{\alpha_d}-E_{m^{c\to d}_{k,l}}
=
(N-k)\big[kU+i\gamma\big].
\label{s:eq:energy_diff_Jx}
\end{align}
Thus the inter-chain hopping amplitude is
\begin{align}
J_x^{(N)}
&= 
\frac12
\left[
\frac{
\prod_{k=0}^{N-1}\langle m^{c\to d}_{k+1,l}|\hat H_{\rm hop}|m^{c\to d}_{k,l}\rangle
}{
\prod_{k=1}^{N-1}(E_{\alpha_c}-E_{m^{c\to d}_{k,l}})
}
+
\frac{ 
\prod_{k=0}^{N-1}\langle m^{c\to d}_{k+1,l}|\hat H_{\rm hop}|m^{c\to d}_{k,l}\rangle
}{
\prod_{k=1}^{N-1}(E_{\alpha_d}-E_{m^{c\to d}_{k,l}})
}
\right]
\nonumber\\
&=
\frac{(-1)^N N J_x^N}{2}
\left[
\prod_{m=1}^{N-1}\frac{1}{mU-i\gamma}
+
\prod_{m=1}^{N-1}\frac{1}{mU+i\gamma}
\right].
\label{s:eq:Jx_final}
\end{align}
Therefore,
\begin{align}
J_0^{(N)}
=
\frac{(-1)^N N J_0^N}{2}
\left[
\prod_{m=1}^{N-1}\frac{1}{mU-i\gamma}
+
\prod_{m=1}^{N-1}\frac{1}{mU+i\gamma}
\right],
\label{s:eq:J0_final}
\end{align}
and
\begin{align}
J_1^{(N)}
=
\frac{(-1)^N N J_1^N}{2}
\left[
\prod_{m=1}^{N-1}\frac{1}{mU-i\gamma}
+
\prod_{m=1}^{N-1}\frac{1}{mU+i\gamma}
\right].
\label{s:eq:J1_final}
\end{align}
The symmetric form guarantees that $J_0^{(N)}$ and $J_1^{(N)}$ are real.

\paragraph{On-site bound-state energies.}
The unperturbed onsite energies are
\begin{align}
E_{\alpha_c}^{(0)}=\frac{U}{2}N(N-1)-iN\gamma,
\qquad
E_{\alpha_d}^{(0)}=\frac{U}{2}N(N-1).
\label{s:eq:E0_bound}
\end{align}
{The hopping terms contribute to the onsite energies via second-order virtual processes, where a particle hops out and back to a lattice site through the same hopping term twice.}
In particular, each of $J$ and $J_1$ provides two such processes toward left and right, respectively, and $J_0$ provides only a single second-order process as it describes a local hopping between the two sublattices within the same position $l$.
Summing all second-order virtual processes gives
\begin{align}
U_c^{(N)}
=
\frac{U}{2}N(N-1)-iN\gamma
+\frac{2N J^2}{(N-1)U}
+\frac{N J_0^2+2N J_1^2}{(N-1)U-i\gamma},
\label{s:eq:UcN_final}
\end{align}
and
\begin{align}
U_d^{(N)}
=
\frac{U}{2}N(N-1)
+\frac{2N J^2}{(N-1)U}
+\frac{N J_0^2+2N J_1^2}{(N-1)U+i\gamma}.
\label{s:eq:UdN_final}
\end{align}
Thus we obtain  the projected $N$-particle bound-state Hamiltonian as
\begin{align}
\hat{H}_{\rm proj}^{(N)}
=&
\Bigg\{
\sum_{l=1}^{L}
\left[
\sum_{\beta=c,d}
J_{\Phi}^{(N)} \hat{\alpha}_{\beta,l}^{\dagger}\hat{\alpha}_{\beta,l+1}
+J_1^{(N)}
\left(
\hat{\alpha}^{\dagger}_{c,l} \hat{\alpha}_{d,l+1}
+\hat{\alpha}^{\dagger}_{d,l} \hat{\alpha}_{c,l+1}
\right)
\right]
+\sum_{l=1}^{L}J_0^{(N)}
\hat{\alpha}^{\dagger}_{c,l} \hat{\alpha}_{d,l}
\Bigg\}+\mathrm{h.c.}\nonumber\\
&+\sum_{l=1}^{L}U_{c}^{(N)} \hat{\alpha}_{c,l}^{\dagger} \hat{\alpha}_{c,l}
+\sum_{l=1}^{L}U_{d}^{(N)} \hat{\alpha}_{d,l}^{\dagger} \hat{\alpha}_{d,l},
\label{s:eq:H_eff_project_N}
\end{align}
with $L+1\equiv1$.

\begin{figure} 
\centering
\includegraphics[width=1
\linewidth]{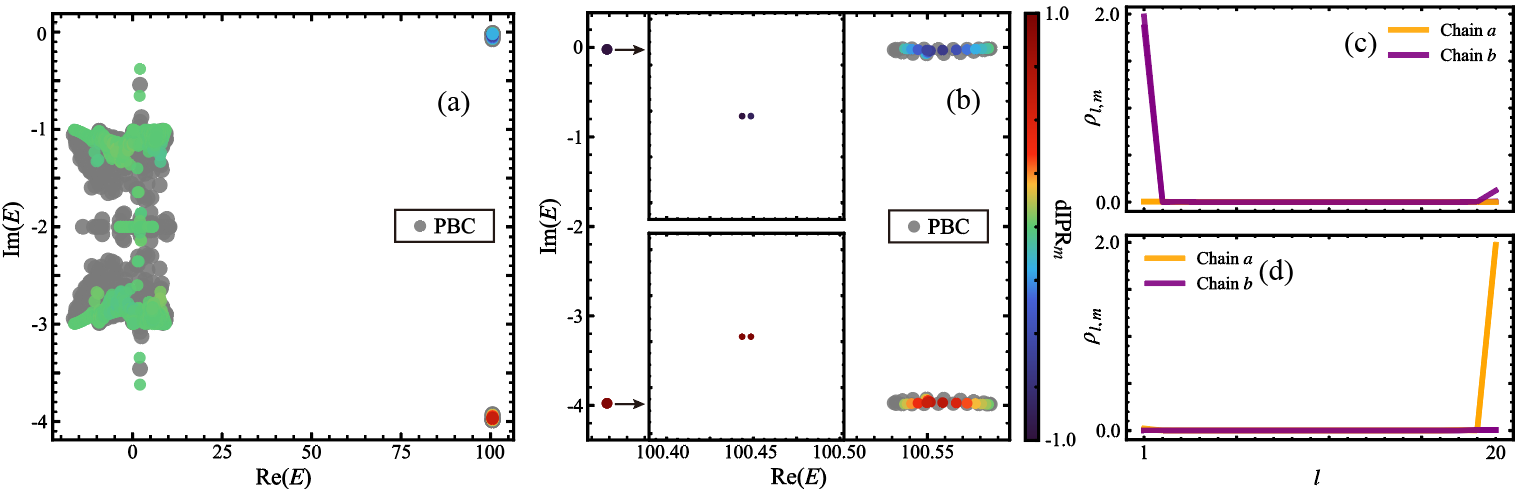}  
\caption{Tamm--Shockley boundary states near the two-particle bound-state spectrum.
(a) Full complex spectrum of $\hat{H}_{a}$ under PBCs (gray) and OBCs, with OBC eigenstates colored by ${\rm dIPR}_m$.
(b) Magnified bound-state sector spectral. The arrows mark two pairs of isolated OBC Tamm--Shockley states, and the insets zoom in on these states.
(c),(d) Particle-number density distributions of representative Tamm--Shockley states selected from the upper and lower zoomed pairs in (b), respectively.
}
\label{s:fig:ts_states}
\end{figure}

\phantomsection
\subsubsection*{C. Tamm--Shockley edge states under OBCs}\label{s:Tamm-Shockley Edge States}
The above effective Hamiltonian assumes PBCs for the system.
On the other hand, we note that for an OBC system, the onsite terms of Eqs.~\eqref{s:eq:UcN_final} and \eqref{s:eq:UdN_final} take different forms on the open boundaries ($l=1$ or $l=L$), where the nonlocal hopping terms $J$ and $J_1$ only connect different sites toward one direction. 
Consequently, the onsite bound-state energies become
\begin{align}
U_{c,(1,L)}^{(N)}
=
\frac{U}{2}N(N-1)-iN\gamma
+\frac{N J^2}{(N-1)U}
+\frac{N J_0^2+N J_1^2}{(N-1)U-i\gamma},
\label{s:eq:Uc_boundary_final}
\end{align}
\begin{align}
U_{d,(1,L)}^{(N)}
=
\frac{U}{2}N(N-1)
+\frac{N J^2}{(N-1)U}
+\frac{N J_0^2+N J_1^2}{(N-1)U+i\gamma}.
\label{s:eq:Ud_boundary_final}
\end{align}
In other words, the interaction effectively induces boundary impurities to the system under OBCs, 
and may give rise to topologically trivial impurity states, known as the Tamm--Shockley edge states~\cite{Tamm1932,Shockley1939},
as shown in Fig.~\ref{s:fig:ts_states}.

\phantomsection
\subsection*{II. Resolved-regime criterion and its breakdown at small \texorpdfstring{\(U\)}{U} and small \texorpdfstring{\(\gamma\)}{gamma}}
\label{s:sec:bound_continuum_hybridization}
The main text focuses on the spectral topology in the resolved parameter regime, where the bound-state clusters are well-separated from each other and other states.
The complementary unresolved regions arise from two distinct mechanisms. 
At small \(U\), the continuum-state sector hybridizes with the bound-state sector,
invalidating the bound-state projection used to derive the effective single-particle Hamiltonian. 
On the other hand, at small \(\gamma\), 
bound states on the two chains are mixed in their imaginary energies, invalidating the decoupled-chain perturbation discussed in Sec. IV.
In this section, we first specify the numerical criterion used to identify the
resolved regime, and then discuss two representative unresolved examples.

 
\begin{figure*}[t]
\centering
\includegraphics[width=0.8\textwidth]{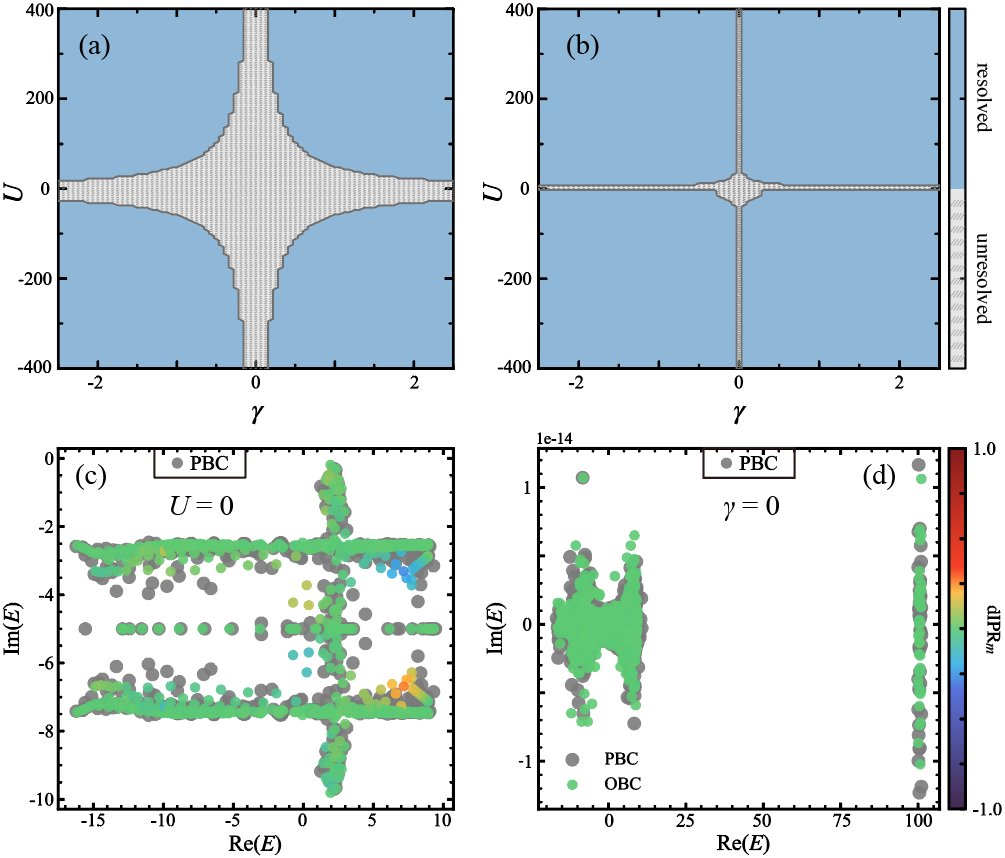}
\caption{   
Resolved-regime masks and representative unresolved spectra.
(a),(b) Resolved regions in the $(\gamma,U)$ plane for
(a) $N=2$, $\theta=\pi/2$, and
(b) $N=3$, $\theta=\pi/6$.
Blue regions satisfy both the bound-continuum separation and imaginary-energy cluster-separation criteria, whereas gray regions fail at least one of them.
(c) At $U=0$ and $\gamma=5$, the bound-state sector is mixed with the scattering continuum, so finite signed ${\rm dIPR}_m$ does not indicate a cluster-resolved bound-state NHSE.
(d) At $U=100$ and $\gamma=0$, the Hamiltonian is Hermitian and the two chain-dominated bound-state clusters are unresolved in imaginary energy; the residual imaginary scale is numerical roundoff.
Gray points denote PBC spectra, and colored points denote OBC eigenstates weighted by ${\rm dIPR}_m$.
Unless otherwise specified, $L=20$, $J=1$, and $J_0=J_1=3$.
}
\label{s:fig:U0_continuum}
\end{figure*}

To distinguish the bound-state sector from the scattering-state continuum, we
define
\begin{align}
C_m=
\bra{\psi_m^R}
\sum_{\beta=a,b}\sum_{l=1}^{L}\hat n_{\beta,l}^{\,2}
\ket{\psi_m^R},
\label{s:eq:Cm_definition}
\end{align}
where \(\ket{\psi_m^R}\) is the \(m\)-th right eigenstate under OBCs. We sort the
eigenstates in descending order of \(C_m\),
\(C_{(1)}\ge C_{(2)}\ge\cdots\), and define the separation between the
bound-state sector and the remaining scattering states as
\(\Delta C_{(2L)}=C_{(2L)}-C_{(2L+1)}\). 
The bound-state sector and the scattering continuum are regarded as fully resolved when
\begin{align}
\Delta C_{(2L)}
\ge
2\max\{\Delta C_{(2L-1)},\,\Delta C_{(2L+1)}\}.
\label{s:eq:Cm_gap_condition}
\end{align}
Once this condition is satisfied, we further require the two bound-state
clusters to be resolved along the imaginary-energy. Specifically, we
sort the imaginary parts of the \(2L\) selected bound-state eigenenergies, 
$y_m = \mathrm{Im}\,E_m $, 
as \(y_1\ge y_2\ge\cdots\ge y_{2L}\), and define
\(g_m=y_m-y_{m+1}\) for \(m=1,2,\ldots,2L-1\). The two clusters are considered
resolved when
\begin{align}
y_L&\ge -\frac{N\gamma}{2}\ge y_{L+1},
\qquad
g_L\ge 2\max_{m\ne L}g_m .
\label{s:eq:strict_gap}  
\end{align}
Here $-\frac{N\gamma}{2}$ is the average imaginary energy for all eigenstates, as each particle acquires on chain $a$ ($b$) an imaginary energy $i\gamma$ ($0$).
We refer to the parameter region satisfying both
Eqs.~\eqref{s:eq:Cm_gap_condition} and~\eqref{s:eq:strict_gap} as the
resolved regime discussed in the main text; parameter points that fail either
condition are classified as unresolved.
Figs.~\ref{s:fig:U0_continuum}(a) and~\ref{s:fig:U0_continuum}(b) show the
resolved/unresolved phase diagrams for \(N=2\) and \(N=3\), respectively. 
The blue regions denote the resolved regime analyzed in the main text, whereas
the hatched regions denote unresolved regimes, where the perturbative
bound-state theory is no longer applicable to the full many-body spectrum.

We now discuss two representative unresolved regimes. At small \(U\) (or \(U=0\)), the
eigenstates in the would-be bound-state clusters are no longer composed
entirely of linear combinations of \(N\)-particle bound-state basis states. 
Instead, bound and scattering basis components strongly mix: scattering eigenstates
acquire weight in the $N$-particle bound-state basis, 
while the nominal bound-cluster states acquire extended
scattering components. Hence the spectrum in
Fig.~\ref{s:fig:U0_continuum}(c) can still display finite signed
\({\rm dIPR}_m\), because any eigenstate with appreciable bound-state components 
can involve virtual hopping processes that carry the statistics-induced effective phase. 

By contrast, Fig.~\ref{s:fig:U0_continuum}(d) shows a different unresolved
limit. Here \(U=100\) keeps the interaction-induced bound sector well separated
from the scattering continuum in real energy, 
but \(\gamma=0\) removes the detuning 
that separates the two bound-state clusters in imaginary energy. 
Moreover, for real hopping amplitudes and interaction strength, 
the Hamiltonian becomes Hermitian at \(\gamma=0\). 
The PBC spectrum has no point-gap, and the bound-state NHSE is absent. 
The quantities \(\overline{\mathrm{dIPR}}\) and
\(\kappa_\beta^{(N)}\) cease to be meaningful because the two
imaginary-energy clusters are not resolved; at the same time, the
statistics-induced bound-state skin effect is absent.

\phantomsection
\subsection*{III. Bloch Hamiltonian and loop criterion}
\label{s:bloch_form_N}

\phantomsection
\subsubsection*{A. Bloch form and branch criterion}\label{s:sec:loop_branch_criterion}

In this subsection, we derive the Bloch form of the projected
$N$-particle bound-state Hamiltonian and use it to analyze the loop structure of the bound-state eigenenergies.
 By performing the Fourier
transformation
\begin{align}
\hat{\alpha}_{\beta,j}
&=
\frac{1}{\sqrt{L}}
\sum_{k\in{\rm BZ}}
e^{ikj}\hat{\alpha}_{\beta,k},
\qquad
\hat{\alpha}_{\beta,j}^{\dagger}
=
\frac{1}{\sqrt{L}}
\sum_{k\in{\rm BZ}}
e^{-ikj}\hat{\alpha}_{\beta,k}^{\dagger}
\end{align}
on Eq.~\eqref{s:eq:H_eff_project_N}, we obtain
\begin{align}
h^{(N)}(k)
=
\begin{pmatrix}
h_{cc}^{(N)}(k) & T^{(N)}(k) \\
T^{(N)}(k) & h_{dd}^{(N)}(k)
\end{pmatrix},
\label{s:Bloch_form_H_eff_N}
\end{align}
with
\begin{align}
h_{cc}^{(N)}(k)
&=
U_c^{(N)}
+
2J^{(N)}
\cos(k-\Phi_N),
\\
h_{dd}^{(N)}(k)
&=
U_d^{(N)}
+
2J^{(N)}
\cos(k-\Phi_N),
\\
T^{(N)}(k)
&=
J_0^{(N)}+2J_1^{(N)}\cos k .
\end{align}
Equivalently,
\begin{align}
h^{(N)}(k)
=
h_0^{(N)}(k)\hat{\sigma}_0
+
\sum_{s=x,y,z}
h_s^{(N)}(k)\hat{\sigma}_s,
\label{s:eq:hk_N}
\end{align}
where the explicit forms of the Hamiltonian, expressed in terms of system parameters, are given by
\begin{align}
h_0^{(N)}(k)
&=
C_0^{(N)}
+
2J^{(N)}
\cos(k-\Phi_N),
\\
h_x^{(N)}(k)
&=
J_0^{(N)}+2J_1^{(N)}\cos k,
\\
h_y^{(N)}(k)
&=0,
\\
h_z^{(N)}(k)
&=
C_z^{(N)},\\
C_0^{(N)}
=
\frac{U_c^{(N)}+U_d^{(N)}}{2}&,
\qquad
C_z^{(N)}
=
\frac{U_c^{(N)}-U_d^{(N)}}{2}.
\end{align}
Note that in Eq.~8 and related discussion of the main text, we have omitted the superscript ``$(N)$” in the Hamiltonian terms $h_{0,x,y,z}^{(N)}$ for simplicity. 
For real hopping amplitudes and loss, $C_z^{(N)}$ takes imaginary values, given by
\begin{align}
C_z^{(N)}
=
\frac{i}{2}
\left[
-N\gamma
+
\frac{2\gamma\left(NJ_0^2+2NJ_1^2\right)}
{(N-1)^2U^2+\gamma^2}
\right].\label{eq.Cz}
\end{align}
The eigenenergies are
\begin{align}
E_\pm^{(N)}(k)
=
h_0^{(N)}(k)
\pm
\Delta E(k),
\label{s:eq:eigenvalues_N}
\end{align}
with
$\Delta E(k)
=
\sqrt{
\bigl[h_x^{(N)}(k)\bigr]^2
+
\bigl[C_z^{(N)}\bigr]^2
}$.

We first derive an analytical criterion for the formation of an individual-band
loop spectrum. For a fixed band index $\eta=\pm$, the eigenenergy in
Eq.~\eqref{s:eq:eigenvalues_N} can be written as
\begin{align}
E_\eta^{(N)}(k)
=
C_0^{(N)}
+
2J^{(N)}\cos(k-\Phi_N)
+
\eta \Delta E(k).
\label{s:eq:E_eta_loop}
\end{align}
The two halves of the Brillouin zone can be resolved by introducing
\begin{align}
x=\cos k,\qquad
\sin k=\sigma\sqrt{1-x^2},
\qquad
\sigma=\pm .
\end{align}
Then the same band $E_\eta^{(N)}(k)$ is decomposed into two branches,
\begin{align}
E_{\eta,\sigma}^{(N)}(x)
=&\,
C_0^{(N)}
+
2J^{(N)} x\cos\Phi_N
+
\eta \Delta E(x)
\nonumber\\
&+
\sigma 2J^{(N)}\sin\Phi_N\sqrt{1-x^2},
\qquad x\in[-1,1].
\label{s:eq:E_eta_branch_loop}
\end{align}
The two branches meet at $x=\pm1$, where $\sqrt{1-x^2}=0$, and together form
the closed trajectory of a single band in the complex-energy plane.

Eq.~\eqref{s:eq:E_eta_branch_loop} makes the loop-forming mechanism
transparent. For a fixed band $\eta$, the two branches have the same imaginary
coordinate,
\begin{align}
{\rm Im}\,E_{\eta,+}^{(N)}(x)
=
{\rm Im}\,E_{\eta,-}^{(N)}(x)
=
{\rm Im}\,C_0^{(N)}
+
\eta\,{\rm Im}\,\Delta E(x),
\label{s:eq:branch_common_imag}
\end{align}
but are separated along the real-energy direction by
\begin{align}
E_{\eta,+}^{(N)}(x)-E_{\eta,-}^{(N)}(x)
=
4J^{(N)}\sin\Phi_N\sqrt{1-x^2}.
\label{s:eq:branch_splitting}
\end{align}
Therefore, an individual band forms a loop only when two ingredients are
simultaneously present:
\begin{align}
J^{(N)}\sin\Phi_N\neq0,
\qquad
{\rm Im}\,\Delta E(x)\not\equiv {\rm const}.
\label{s:eq:loop_criterion}
\end{align}
The first condition separates the two branches with $\sigma=\pm1$, while the second provides a
nonconstant imaginary-energy coordinate. 
For $J^{(N)}\sin\Phi_N=0$,
the two branches coincide
and the spectrum overlaps with itself at $k$ and $-k$ [i.e., $E^{(N)}_\eta(k)=E^{(N)}_\eta(-k)$].
If the second condition is violated, the two branches 
remain at a fixed imaginary energy and cannot
enclose a finite area in the complex plane. Thus Eq.~\eqref{s:eq:loop_criterion}
gives the criterion for forming a spectral loop.

\phantomsection
\subsubsection*{B. Signed spectral area}\label{s:sec:signed_area}
To characterize the spectral loop, we now derive the signed spectral area used in the main text. This quantity is not
a topological invariant, but it provides a compact analytical measure of the
orientation and parameter dependence of the spectral loop.
For each band $\eta=\pm$, we write
\begin{align}
E_\eta^{(N)}(k)
=
X_\eta^{(N)}(k)+iY_\eta^{(N)}(k),
\end{align}
where
\begin{align}
X_\eta^{(N)}(k)
&=
c_{0r}^{(N)}
+
2J^{(N)}\cos(k-\Phi_N)
+
\eta u_N(k),
\label{s:eq:X_eta_def}
\\
Y_\eta^{(N)}(k)
&=
c_{0i}^{(N)}
+
\eta v_N(k).
\label{s:eq:Y_eta_def}
\end{align}
Here
\begin{align}
C_0^{(N)}=c_{0r}^{(N)}+ic_{0i}^{(N)},
\qquad
\Delta E(k)=u_N(k)+iv_N(k).
\end{align}
Since
$h_x^{(N)}(k)=J_0^{(N)}+2J_1^{(N)}\cos k$ is even in $k$, both $u_N(k)$ and
$v_N(k)$ are even functions of $k$, while their derivatives are odd.

The signed area enclosed by the parametric curve
$\bigl(X_\eta^{(N)}(k),Y_\eta^{(N)}(k)\bigr)$ is
\begin{align}
S_\eta^{(N)}
=
\frac{1}{2}
\int_{-\pi}^{\pi}
\left[
X_\eta^{(N)}(k)
\frac{dY_\eta^{(N)}(k)}{dk}
-
Y_\eta^{(N)}(k)
\frac{dX_\eta^{(N)}(k)}{dk}
\right]dk .
\label{s:eq:signed_area_definition}
\end{align}
Substituting Eqs.~\eqref{s:eq:X_eta_def} and
\eqref{s:eq:Y_eta_def}, the terms involving $c_{0r}^{(N)}$,
$c_{0i}^{(N)}$, and $u_N(k)$ vanish after integration because of periodicity
and parity. The remaining contribution is
\begin{align}
S_\eta^{(N)}
=
\eta J^{(N)}
\int_{-\pi}^{\pi}
\left[
v_N'(k)\cos(k-\Phi_N)
+
v_N(k)\sin(k-\Phi_N)
\right]dk .
\label{s:eq:signed_area_intermediate}
\end{align}
Using integration by parts,
\begin{align}
\int_{-\pi}^{\pi}
v_N'(k)\cos(k-\Phi_N)\,dk
=
\int_{-\pi}^{\pi}
v_N(k)\sin(k-\Phi_N)\,dk ,
\end{align}
where the boundary term vanishes by periodicity, we obtain
\begin{align}
S_\eta^{(N)}
=
2\eta J^{(N)}
\int_{-\pi}^{\pi}
v_N(k)\sin(k-\Phi_N)\,dk .
\label{s:eq:signed_area_before_parity}
\end{align}
Finally, using
\begin{align}
\sin(k-\Phi_N)
=
\sin k\cos\Phi_N-\cos k\sin\Phi_N
\end{align}
and the fact that $v_N(k)\sin k$ is odd in $k$, the first term integrates to
zero. Hence
\begin{align}
S_\eta^{(N)}
=
-2\eta J^{(N)}
\sin\Phi_N
\int_{-\pi}^{\pi}
v_N(k)\cos k\,dk ,
\label{s:eq:area_result_final_N}
\end{align}
where
\begin{align}
v_N(k)
=\mathrm{Im} \Delta E(k)=
\mathrm{Im}
\sqrt{
\bigl[J_0^{(N)}+2J_1^{(N)}\cos k\bigr]^2
+
\bigl(C_z^{(N)}\bigr)^2
}.
\label{s:eq:vk_expression_N}
\end{align}

\phantomsection
\subsubsection*{C. Vanishing-area conditions and figure-eight exception}\label{s:sec:area_zero_conditions}
First, we note that $S_\eta^{(N)}=0$ under two conditions: (i) the absence of a spectral loop, or (ii) the spectral loop forms two symmetric sub-loops with opposite winding direction, so that their contributions to the signed area cancel each other. 
Below we list the different parameter conditions that lead to $S_\eta^{(N)}=0$.

\paragraph{$N=1$, $J=0$, or $|U|\to\infty$.}
These conditions ensure a vanishing prefactor $J^{(N)}\sin\Phi_N=0$ in the integral Eq.~\eqref{s:eq:area_result_final_N}.
Specifically, the single-particle limit yields $\Phi_N=0$, and the other two conditions lead to $J^{(N)}=0$, see Eq.~\eqref{s:eq:Jphi_def}.
In addition, from Eq.~\eqref{s:eq:Jphi_def}, we can see that the sign of $J^{(N)}$ depends on the parity of $N$, and also on the sign of $U$ if $N$ takes even values, resulting in the parity-dependence of the orientation of the spectral winding.

As discussed for Eq.~\eqref{s:eq:loop_criterion}, $J^{(N)}\sin\Phi_N=0$ indicates the overlapping between two branches of eigenenergies ($k\in[0,\pi]$ and $k\in[-\pi,0]$), thus they cannot form a nontrivial spectral loop.

\paragraph{Hermitian ($\gamma=0$) and anti-Hermitian limits ($\gamma\rightarrow \infty$).}
Following Eq.~\eqref{eq.Cz}, $\gamma=0$ leads to $C_z^{(N)}=0$ and thus $v_N(k)=0$, and $\gamma\rightarrow \infty$ leads to a constant $vN(k)\approx |C_z^{(N)}|$, both resulting in a vanishing integral in Eq.~\eqref{s:eq:area_result_final_N} and therefore $S_\eta^{(N)}=0$, which also indicates the absence of a spectral loop [see the discussion of Eq.~\eqref{s:eq:loop_criterion}].

\paragraph{No momentum-dependent hybridization, $J_1=0$.}
When $J_1^{(N)}=0$, one has $\Delta E(k)=\sqrt{[J_0^{(N)}]^2+[C_z^{(N)}]^2}$ becoming $k$-independent. Thus the integral in Eq.~\eqref{s:eq:area_result_final_N} vanishes and a nontrivial spectral loop is absent, as in the previous case.

\paragraph{No onsite hybridization, $J_0=0$.}
This case is qualitatively different. Here
\begin{align}
v_N(k)
=
\mathrm{Im}
\sqrt{
4\bigl(J_1^{(N)}\bigr)^2\cos^2 k
+
\bigl(C_z^{(N)}\bigr)^2
},
\end{align}
which is symmetric under $k\rightarrow\pi-k$, whereas $\cos k$ is
antisymmetric under the same transformation. Therefore the integral in
Eq.~\eqref{s:eq:area_result_final_N} vanishes and
$S_\eta^{(N)}=0$. However, this is caused by symmetry-enforced
signed-area cancellation. If
\begin{align}
J^{(N)}\sin\Phi_N\neq0,
\qquad
{\rm Im}\,\Delta E(x)\not\equiv{\rm const},
\end{align}
the two branches in Eq.~\eqref{s:eq:E_eta_branch_loop} still form a
self-intersecting figure-eight-type spectrum. The two lobes carry opposite
orientations, so their signed areas cancel. Therefore $J_0=0$ should not be
interpreted as a general absence of point-gap winding or NHSE.

\begin{figure} 
\centering
\includegraphics[width=0.8\linewidth]{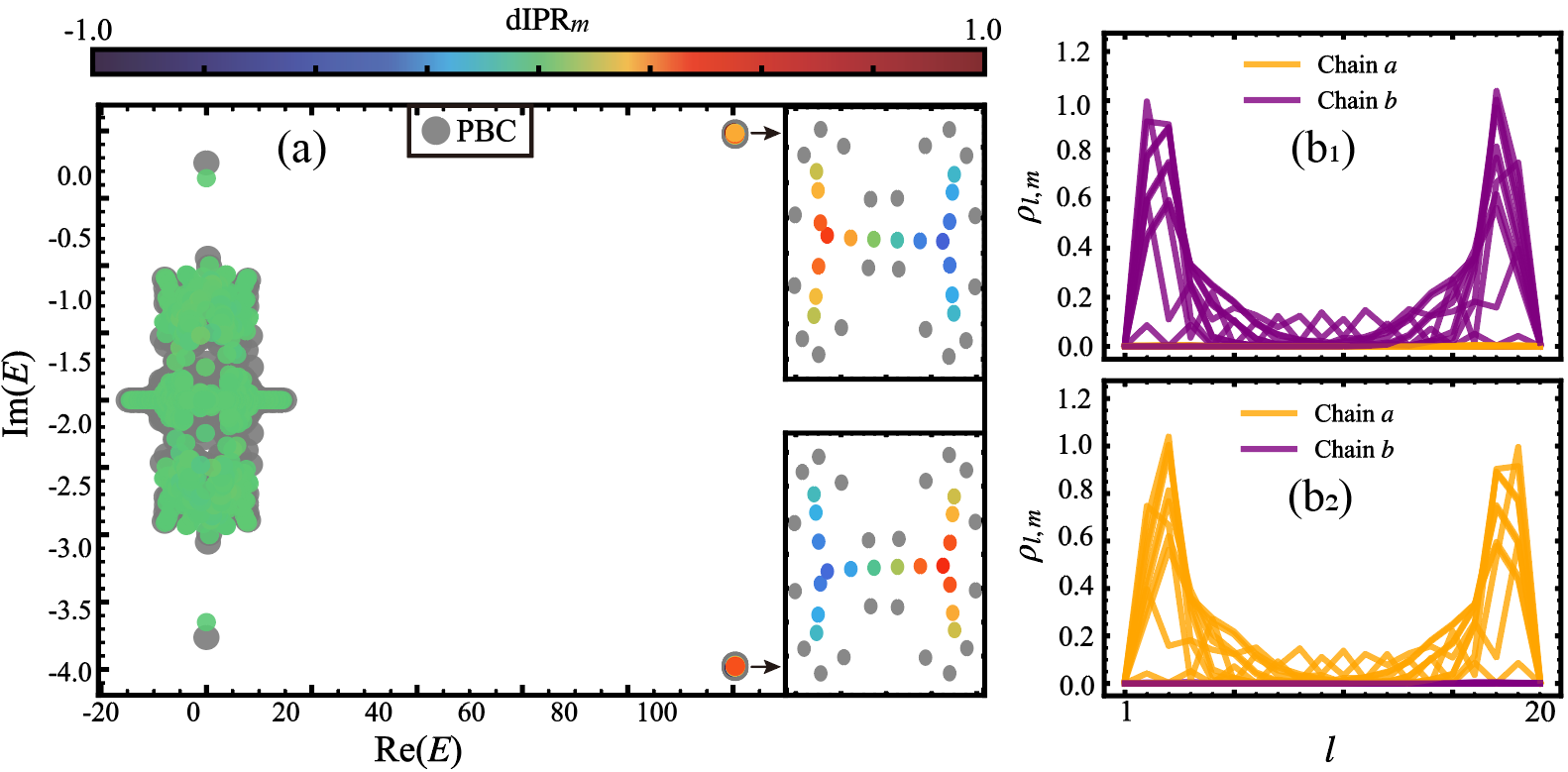}  
\caption{
(a) Complex spectrum for \(N=2\), \(J_0=0\), and \(\theta=\pi/4\). PBC
eigenvalues are shown in gray, while OBC eigenstates are colored by
\({\rm dIPR}_m\). The insets magnify the two bound-state clusters near
\({\rm Re}(E)\simeq U\). The vanishing signed area at \(J_0=0\) results from a
symmetric figure-eight-type trajectory rather than loop vanishing.
(b1),(b2) Chain-resolved OBC density profiles of the two magnified clusters in
(a), showing boundary accumulation despite \(S_\pm=0\).
Other parameters are $L=20$, $J_1=3$, $J=1$, $\gamma=2$, and $U=100$.
}
\label{s:fig:area_exception_parity}
\end{figure}

Figs.~\ref{s:fig:area_exception_parity}(a) and ~\ref{s:fig:area_exception_parity}(b) illustrate
such an example. The net signed area vanishes because the two lobes of a self-intersecting spectrum
carry opposite orientations, rather than because the point-gap structure
collapses. 

\phantomsection
\subsection*{IV. Effective non-reciprocity from decoupled-chain perturbation}\label{s:sec:decoupled_perturbation_theory_N}
In our numerical results, we observe that the two bound-state clusters separated by a strong dissipation $\gamma$ have eigenstates occupying mostly on one of the two chains, suggesting an effectively decoupled scenario between them.
Therefore, in this subsection, we employ a perturbative approach to approximate the $N$-particle bound-state Hamiltonian as two effectively decoupled chains. 


As derived in the previous subsection, Eq.~\eqref{s:Bloch_form_H_eff_N} yields the Bloch form of the $N$-particle single-particle effective Hamiltonian:
\begin{align}\label{s:Bloch_form_H_eff_N_pert}
h^{(N)}(k)=
\begin{pmatrix}
h_{cc}^{(N)}(k) & T^{(N)}(k) \\
T^{(N)}(k) & h_{dd}^{(N)}(k)
\end{pmatrix}.
\end{align}
We treat the off-diagonal term $T^{(N)}(k)$ (originating from the inter-chain couplings $J_0^{(N)}$ and $J_1^{(N)}$) as perturbation to the diagonal terms $h_{cc}^{(N)}(k)$ and $h_{dd}^{(N)}(k)$. The Hamiltonian is thus partitioned as:
\begin{align}
    h^{(N)}(k) = h_{\rm diag}^{(N)}(k) + V^{(N)}(k) = 
    \begin{pmatrix}
    h_{cc}^{(N)}(k) & 0\\
    0 & h_{dd}^{(N)}(k)
    \end{pmatrix} +
    \begin{pmatrix}
    0 & T^{(N)}(k)\\
    T^{(N)}(k) & 0
    \end{pmatrix}.
\end{align}
We perform a similarity transformation on $h^{(N)}(k)$ to decouple the chains up to leading order:
\begin{align}
    \tilde h^{(N)}(k) &\equiv e^{S(k)} h^{(N)}(k) e^{-S(k)} \nonumber\\
    &= h_{\rm diag}^{(N)}(k) + V^{(N)}(k)
    + [S(k),h_{\rm diag}^{(N)}(k)] + [S(k),V^{(N)}(k)]
    + \frac{1}{2}[S(k),[S(k),h_{\rm diag}^{(N)}(k)]]
    + \mathcal O(S^3(k))\label{s:H_bch_N}
\end{align}
where the transformation matrix is parametrized as
\begin{align}
S(k) =
\begin{pmatrix}
0 & X_1(k)\\
-\,X_2(k) & 0
\end{pmatrix}.
\end{align}
To eliminate the first-order off-diagonal term, we impose the condition:
\begin{align}
    V^{(N)}(k) + [S(k),h_{\rm diag}^{(N)}(k)] = 0.\label{s:eliminate_condition_N}
\end{align}
This leads to the Sylvester equations, yielding the scalar solutions:
\begin{align}
X_1(k)&=X_2(k)=\frac{T^{(N)}(k)}{2C_z^{(N)}},
\label{s:X_Y_N}
\end{align}
Substituting Eq.~\eqref{s:eliminate_condition_N} into Eq.~\eqref{s:H_bch_N}, the remaining second-order correction is $\frac{1}{2}[S(k),V^{(N)}(k)]$. The decoupled effective Hamiltonian reads:
\begin{align}
    \tilde h^{(N)}(k) \approx
h_{\rm diag}^{(N)}(k)  + \frac{1}{2}[S(k),V^{(N)}(k)]=
    \begin{pmatrix}
    h_{cc}^{(N)}(k) + \frac{(T^{(N)}(k))^2}{2C_z^{(N)}} & 0\\
    0 & h_{dd}^{(N)}(k) - \frac{(T^{(N)}(k))^2}{2C_z^{(N)}}
    \end{pmatrix} \label{s:H_eff_N},
\end{align}
where the squared perturbation expands as
\begin{align}
\bigl(T^{(N)}(k)\bigr)^2
=\bigl(J_0^{(N)}\bigr)^2+4J_0^{(N)}J_1^{(N)}\cos k + 2\bigl(J_1^{(N)}\bigr)^2\bigl(1+\cos 2k\bigr).
\end{align}
By transforming back to real space, we obtain the effective Hamiltonian for the $c$-chain:
\begin{align}
H_{\mathrm{eff}}^{(c,N)}
&=\sum_{l}\varepsilon_c^{\mathrm{eff},(N)}\ \hat\alpha_{c,l}^\dagger\hat\alpha_{c,l}
+\sum_{l}\Big(t_{c,R}^{\mathrm{eff},(N)}\ \hat\alpha_{c,l+1}^\dagger\hat\alpha_{c,l}
+t_{c,L}^{\mathrm{eff},(N)}\ \hat\alpha_{c,l}^\dagger\hat\alpha_{c,l+1}\Big)\nonumber\\
&\quad+\sum_{l}\Big(t_{c,2}^{\mathrm{eff},(N)}\ \hat\alpha_{c,l+2}^\dagger\hat\alpha_{c,l}
+t_{c,2}^{\mathrm{eff},(N)}\ \hat\alpha_{c,l}^\dagger\hat\alpha_{c,l+2}\Big),
\end{align}
with the renormalized coefficients:
\begin{align}
\varepsilon_c^{\mathrm{eff},(N)}&=U_c^{(N)}+\frac{(J_0^{(N)})^2+2(J_1^{(N)})^2}{2C_z^{(N)}}, \qquad t_{c,2}^{\mathrm{eff},(N)}=\frac{(J_1^{(N)})^2}{2C_z^{(N)}},\\
t_{c,R}^{\mathrm{eff},(N)}&=J^{(N)} e^{+i\Phi_N}-iY^{(N)},\qquad
t_{c,L}^{\mathrm{eff},(N)}=J^{(N)} e^{-i\Phi_N}-iY^{(N)}.\label{s:t_eff_c_N}
\end{align}
Similarly, for the $d$-chain:
\begin{align}
H_{\mathrm{eff}}^{(d,N)}
&=\sum_{l}\varepsilon_d^{\mathrm{eff},(N)}\ \hat\alpha_{d,l}^\dagger\hat\alpha_{d,l}
+\sum_{l}\Big(t_{d,R}^{\mathrm{eff},(N)}\ \hat\alpha_{d,l+1}^\dagger\hat\alpha_{d,l}
+t_{d,L}^{\mathrm{eff},(N)}\ \hat\alpha_{d,l}^\dagger\hat\alpha_{d,l+1}\Big)\nonumber\\
&\quad+\sum_{l}\Big(t_{d,2}^{\mathrm{eff},(N)}\ \hat\alpha_{d,l+2}^\dagger\hat\alpha_{d,l}
+t_{d,2}^{\mathrm{eff},(N)}\ \hat\alpha_{d,l}^\dagger\hat\alpha_{d,l+2}\Big),
\end{align}
with the renormalized coefficients:
\begin{align}
\varepsilon_d^{\mathrm{eff},(N)}
&=
U_d^{(N)}
-
\frac{(J_0^{(N)})^2+2(J_1^{(N)})^2}{2C_z^{(N)}},
\qquad
t_{d,2}^{\mathrm{eff},(N)}
=
-\frac{(J_1^{(N)})^2}{2C_z^{(N)}},
\\
t_{d,R}^{\mathrm{eff},(N)}
&=
J^{(N)} e^{+i\Phi_N}
+iY^{(N)},
\qquad
t_{d,L}^{\mathrm{eff},(N)}
=
J^{(N)} e^{-i\Phi_N}
+iY^{(N)}.
\label{s:t_eff_d_N}
\end{align}
with $Y^{(N)}=\frac{i J_0^{(N)} J_1^{(N)}}{C_z^{(N)}}$ being a purely real coefficient [see Eqs.~\eqref{s:eq:J0_final},~\eqref{s:eq:J1_final}, and~\eqref{eq.Cz}].

In the above effective models, non-reciprocity arises from $Y^{(N)}$ and the imaginary part of $J^{(N)} e^{\pm i\Phi_N}$,
while the second-neighbor hopping \(t_{\beta,2}^{\mathrm{eff},(N)}\) ($\beta=c,d$) is reciprocal.
More importantly,
it is comparably much weaker than the nearest-neighbor hopping in the strong-interaction limit.
Explicitly,  
\begin{align}
|Y^{(N)}|\simeq|2t_{\beta,2}^{\mathrm{eff},(N)}|,\qquad
\frac{
\left|t_{\beta,2}^{\mathrm{eff},(N)}\right|
}{
\left|J^{(N)}\right|
}
\simeq
\frac{1}{(N-1)!}\,
\frac{\left|J_1\right|^{2N}}
{\left|\gamma\right|\left|J\right|^N\left|U\right|^{N-1}} \ll1,
\label{s:eq:t2_scale_N}
\end{align}
when $U\rightarrow \infty$ (which also gives  \(C_z^{(N)}\simeq -iN\gamma/2\) ) and the hopping amplitudes are of the same order of magnitude [$J_0^{(N)}\simeq J_1^{(N)}$].
Therefore, we characterize the non-reciprocity through
\begin{align}
\kappa_{\beta}^{(N)}
=
\frac{1}{2}
\ln
\frac{
|t_{\beta,R}^{\mathrm{eff},(N)}|
}{
|t_{\beta,L}^{\mathrm{eff},(N)}|
},
\qquad
\beta\in\{c,d\}.
\end{align}
This quantity reduces to the inverse localization length of skin modes in the effective
single-chain limit where the symmetric next-nearest-neighbor hopping and
residual interchain mixing can be omitted. More generally, it should be viewed
as a leading diagnostic of the effective-chain non-reciprocity, rather than as a
standalone criterion for the existence or absence of NHSE in the full two-band
problem.

Defining
\begin{align}
\Sigma^{(N)}
&\equiv
\bigl(J^{(N)}\bigr)^2
+
\bigl(Y^{(N)}\bigr)^2,
\\
\Omega_c^{(N)}(\Phi_N)
&\equiv
-2J^{(N)}Y^{(N)}\sin\Phi_N,
\\
\Omega_d^{(N)}(\Phi_N)
&\equiv
+2J^{(N)}Y^{(N)}\sin\Phi_N,
\end{align}
we obtain
\begin{align}
|t_{c,R}^{\mathrm{eff},(N)}|^2
&=
\Sigma^{(N)}+\Omega_c^{(N)}(\Phi_N),
&
|t_{c,L}^{\mathrm{eff},(N)}|^2
&=
\Sigma^{(N)}-\Omega_c^{(N)}(\Phi_N),
\\
|t_{d,R}^{\mathrm{eff},(N)}|^2
&=
\Sigma^{(N)}+\Omega_d^{(N)}(\Phi_N),
&
|t_{d,L}^{\mathrm{eff},(N)}|^2
&=
\Sigma^{(N)}-\Omega_d^{(N)}(\Phi_N).
\end{align}
Thus
\begin{align}
\kappa_c^{(N)}
=
\frac{1}{4}
\ln
\frac{
\bigl(J^{(N)}\bigr)^2+\bigl(Y^{(N)}\bigr)^2
-2J^{(N)}Y^{(N)}\sin\Phi_N
}{
\bigl(J^{(N)}\bigr)^2+\bigl(Y^{(N)}\bigr)^2
+2J^{(N)}Y^{(N)}\sin\Phi_N
},
\label{s:eq:kappa_c_N}
\end{align}
and
\begin{align}
\kappa_d^{(N)}=-\kappa_c^{(N)}.
\label{s:eq:kappa_d_N}
\end{align}
Eq.~\eqref{s:eq:kappa_c_N} and Eq.~\eqref{s:eq:kappa_d_N} show that the
leading effective non-reciprocity is bipolar: the two chains acquire opposite
non-reciprocal pumping directions. This explains the opposite skin
accumulations observed in the simple-loop regimes discussed in the main text.

\bibliography{refs}

\end{document}